%% file: main.tex
\documentclass{article}

\usepackage{arxiv}

\usepackage[utf8]{inputenc} 
\usepackage[T1]{fontenc}    
\usepackage{hyperref}       
\usepackage{url}            
\usepackage{booktabs}       
\usepackage{amsfonts}       
\usepackage{nicefrac}       
\usepackage{microtype}      
\usepackage{lipsum}

\usepackage{cite}
\usepackage{amsmath,amssymb,amsfonts}
\usepackage{algorithmic}
\usepackage{graphicx}
\usepackage{textcomp}
\usepackage{xcolor}

\input{settings}

\title{Multi-core Fiber and Power-limited Optical Network Topology Optimization with MILP}

\author{
  Bjoern Annighoefer \\
  Institute of Aircraft Systems\\
  University of Stuttgart\\
  Stuttgart, Germany \\
	\texttt{0000-0002-1268-0862}
   \And
  Adrian Zeyher \\
  Institute of Aircraft Systems\\
  University of Stuttgart\\
  Stuttgart, Germany \\
  \texttt{0000-0003-2888-4509} \\
  \And
  Johannes Reinhart \\
	Institute of Aircraft Systems\\
  University of Stuttgart\\
  Stuttgart, Germany \\
  \texttt{0000-0002-3512-5220} \\
}

\begin{document}

\twocolumn[

\maketitle

\input{abstract}

\vspace{1cm}
]

\input{introduction}

\input{problem}

\input{related_work}

\input{theory}

\input{validation}

\input{results}

\input{conclusion}

\section*{Acknowledgements}
This paper is based on research work carried out in the DELIA project (contract code: 16KIS0940) funded by the German Federal Ministry of Education and Research (BMWI) in the IKT 2020 program.



\input{main.bbl}

\end{document}

%% file: settings.tex
\usepackage{siunitx}					
\newunit{\dBm}{dBm}

\usepackage{tabu}
\usepackage{xparse}

\usepackage{import}
\usepackage{transparent}

\usepackage{tikz}
\usetikzlibrary{patterns}
\usetikzlibrary{shapes}
\usetikzlibrary{plotmarks}

\usepackage{tikzscale}

\usepackage{multicol}
\usepackage{tabularx}
\usepackage{booktabs}

\usepackage{amsmath}			
\usepackage{amsthm}
\usepackage{commath}
\usepackage{mathtools} 

\allowdisplaybreaks

\usepackage{amsfonts}

\DeclareDocumentCommand{\place}{mgg}{%
	\ensuremath{%
		\IfNoValueTF {#2}%
		{ {#1} }%
		{%
			\IfNoValueTF {#3}%
			{ {#1}^{#2} }%
			{ {#1}^{#2}_{#3}  }%
		}%
	}%
}

\DeclareDocumentCommand{\vec}{smgg}{%
	\ensuremath{%
		\IfBooleanTF {#1}%
			{ \begin{pmatrix}#2\end{pmatrix} }%
			{ \place{\mathbf{#2}}{#3}{#4} }
	}%
}

\DeclareDocumentCommand{\matrix}{smgg}{%
	\ensuremath{%
		\IfBooleanTF {#1}%
			{ \begin{bmatrix}#2\end{bmatrix} }%
			{ \place{\mathbf{#2}}{#3}{#4} }
	}%
}

\DeclareDocumentCommand{\set}{m}{ \ensuremath{\mathcal{#1}} }
\DeclareDocumentCommand{\abs}{m}{ \ensuremath{\left|#1\right|} }
\DeclareDocumentCommand{\sumNorm}{m}{ \ensuremath{\norm{#1}_1 } }

\DeclareDocumentCommand{\transpose}{m}{\ensuremath{#1^{\mkern-1.5mu\mathsf{T}}}}

\DeclareDocumentCommand{\reals}{}{\ensuremath{\mathbb{R}}}
\DeclareDocumentCommand{\integers}{}{\ensuremath{\mathbb{Z}}}
\DeclareDocumentCommand{\binary}{}{\ensuremath{\left\{0,1\right\}}}

\DeclareDocumentCommand{\var}{s O{x} m G{}}{%
	\ensuremath{%
		\IfBooleanTF {#1}%
			{\place{#2}{#3}{ \text{#4} }}%
			{\place{#2}{#3}{ #4 }}%
	}%
}

\DeclareDocumentCommand{\property}{s O{p} m m}{%
	\ensuremath{%
		\IfBooleanTF {#1}%
			{\place{#2}{#3}{\text{#4} }}%
			{\place{#2}{#3}{ #4 }}%
	}%
}

\DeclareDocumentCommand{\binaryNot}{m}{%
	\ensuremath{%
		\left( \num{1} - #1 \right)%
	}%
}

\DeclareDocumentCommand{\bigMexpr}{som}{%
	\ensuremath{%
		\IfNoValueTF {#2}%
			{ P_{lim} }%
			{ {#2} }%
		\cdot%
		\IfBooleanTF {#1}%
			{ #3 }%
			{ \binaryNot{#3} }%
	}%
}

\DeclareDocumentCommand{\underlineText}{m}{\ensuremath{\underline{\text{#1}}}}
\DeclareDocumentCommand{\overlineText}{m}{\ensuremath{\overline{\text{#1}}}}

%% file: abstract.tex
\begin{abstract}
Optical networks with multi-core fibers can replace several electronics networks with a single topology. Each electronic link is replaced by a single fiber, which can save space, weight, and cost, while having better segregation and EMI resistance. This is, for instance, of high interest in safety-critical cyber-physical systems, such as aircraft avionics networks. Finding the optimal topology requires finding the optimal number of components, component locations, inter-meshing, and signal routing, while assuring the appropriate optical power level at each participating device. A Mixed-integer Linear Programming (MILP) representation is presented for the optimization of the topology of optical multi-core fiber networks. The optimization approach retrieves a globally optimal topology with respect to weight or cost, i.e. it builds the optimal network topology from a set of switch and cable types, which differ in the number of fibers, attenuation, connectors, and other properties. The novelty of our approach is the consideration of translucent and opaque optical switches as well as the representation of cable and device attenuation directly in the MILP constraints. Moreover, arbitrary installation and routing resource restrictions are considered. The application of the approach to five dedicated scenarios yields in each case an optimal solution and validates our method. The application to an excerpt of an aircraft cabin network shows the retrieval of the global optimum in less 30 min for a topology with 48 signals and 23 components.


\end{abstract}

%% file: introduction.tex
\section{Introduction}
\label{sec:introduction}
Optical networks differ from electrical networks by using light as an information carrier. 
Instead of voltage changes in metallic conductors, modulated light is passed from one node to another through fine optical fibers, also known as fiber optics or fiber.
This results in much higher bandwidths compared to electrical networks, while the weight and volume of the components is smaller.
Furthermore, optical networks do not generate electromagnetic interference during transmission and do not themselves react to such interference. 
In addition, less electrical power is required to transmit signals, which has an additional positive impact on weight and costs. (cf.  \cite{Chatterjee2017})

No electromagnetic interference as well as weight, power, and space saving are highly desirable properties in the field of aviation. 
Modern large transport airplanes are based on Distributed Integrated Modular Avionics (DIMA). DIMA is a concept for a safety-critical sharing of computing and network resources. 
Throughout the aircraft fuselage, generic computer modules are installed. 
Locally, I/O modules read sensors and control actuators. Centrally placed processing modules calculate control commands. 
All DIMA modules are connected with a common high-bandwidth Aircraft Data and Communication Network (ADCN). 
The modules and network are configurable.  
The planning and development of a DIMA avionics system architecture is time and resource intensive, due to a high number of software components, signals, modules, peripherals as well as a complex interaction. 
It is assumed that potential is wasted and a (partial) automation of design steps results in better reliability, weight, cost, etc. ~\cite[p.~2]{AnnighoeferBuch}. 
Special attention has to be paid to the network connecting the modules. 
Many implemented functions are safety-relevant and have mandatory reliability and latency requirements. 
A prominent example is the flight control system. 
Since the number of functions increases exponentially over time \cite{itier:2009}, the ADCNs used today will reach their limits in terms of data rate and latency in a foreseeable time. 
Optical networks are being considered for future deployments as an alternative to the commonly used ARINC~629 and ARINC~664 (AFDX) networks \cite{PatentAfdxOptical}.

This work presents a Mixed-Integer Linear Programming (MILP) representation, which generates an optimal network topology for a given installation space from a given set of signals and optical components. 
The algorithm considers necessary signal flows, cable properties such as attenuation and the number of fibers, and generic resource constraints of hardware and locations.

The remainder of this work is organized as follows. 
Chapter \ref{sec:problem} defines our topology optimization problem and involved components. 
In Chapter \ref{sec:related_work} we review the state-of-the-art and compare it to our approach.
Chapter \ref{sec:theory} derives the MILP representation of the topology optimization problem. 
In chapter \ref{sec:validation} the algorithm is validate using five dedicated scenarios. 
In chapter \ref{sec:results} the algorithm is applied to optimize the network of an In-flight Entertainment System (IFE). 
Chapter \ref{sec:conclusion} discusses the results and gives an outlook.

%% file: problem.tex
\section{Topology Optimization Problem and Limitations}
\label{sec:problem}

\begin{figure}[htbp]
	\centering
	\fontsize{6}{5}
	\def\svgwidth{\columnwidth}
	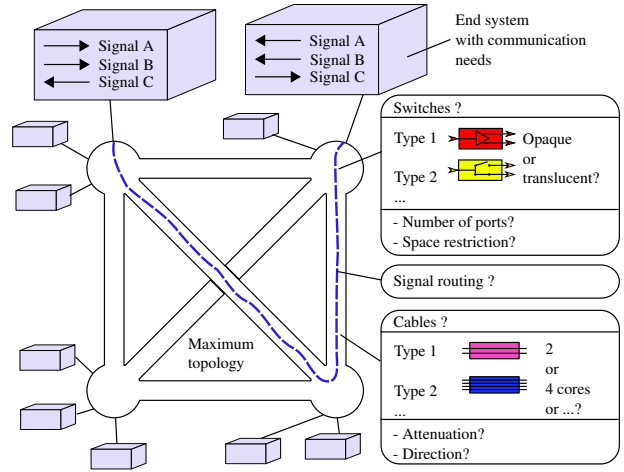
	\caption{Scope of topology optimization}
	\label{fig:problem}
\end{figure}

The general scope of topology optimization in this work is depicted in fig. \ref{fig:problem}. We assume a given installation space for the optical network consisting of possible switch and end system installation locations as well as cable routes. Within this installation space a number of end systems exist that have communications needs in terms of signals, which have to be transmitted and received. Given the installation locations and possible cable routes, an optimal optical network is desired that connects all end systems and fulfills their communication needs. The communication needs are fulfilled if a path from sender to receiver for each signal is determined. The optical network has to be built from predefined switch and cable types. Switches can either be opaque or translucent. Opaque switches convert optical information to electrical and back and, therefore, decouple the optical power of connected cables. Translucent switches forward the original optical signal, i.e. input and output power levels are dependent. Moreover, switches can differ in the number of ports, costs as well as generic resources required for their installation (e.g. space) and internal resources required for signal routing (e.g. bandwidth or number of signals). Cables differ in cost, weight, and resources like switches. Moreover, multi-core fiber cables are assumed. Different cable types can offer a different number of cores. It is assumed that each signal requires a single core. 
This allows for a physical segregation of signals. A valid topology must not exceed the resources of any installation space or cable route. A valid routing must not exceed the resources of any switch or cable and the optical power level of signals has to be in the min-max-range acceptable for the individual components. Attenuation shall be considered in cables, connectors, and translucent devices. Multiple optimization objectives exist, e.g. the cost or weight of the network components, the installation effort, and the number of components in total. A pre-definition of single aspects shall be possible, e.g. fixed switches or cables.

\subsection{Multi-core Fiber Transmission}
\label{sec:opticalnetworks:multicore}

By using multiple cores in a single optical cable, the transmission speed can be increased with the help of parallel transmission.
We refers to cores instead of fibers to prevent the ambiguity of fiber as a synonym for optical cables.
The use of a single cable with multiple cores is common, which are called Multi-Core Fibers (MCF).

\begin{figure}[h!t]
	\centering
	\def\svgwidth{0.75\columnwidth}
	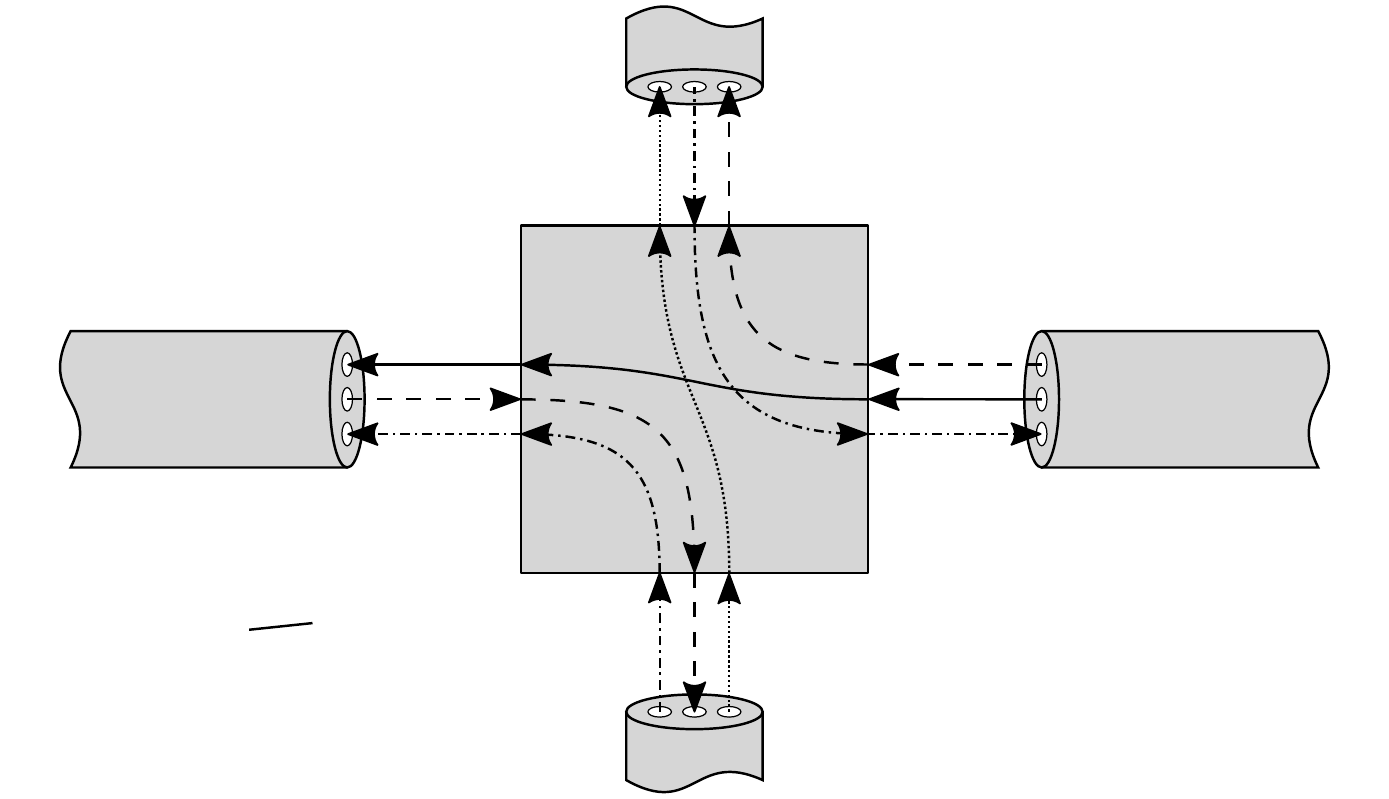
	\caption{Signal path within an optical network that uses multi-core fiber transmission.}
	\label{fig:mcf:switch}
\end{figure}

Fig. \ref{fig:mcf:switch} shows an example of a signal path in a switch. Each of the connected cables has three cores. Each signal is assigned to its own core. The routing of which signal passes through which core can be done dynamically or statically. In the latter case, the paths are calculated in advance as suggested in this work. The physical routing method depends on the switch type, i.e. how the signal is transferred from one cable to the other. 

Due to physical coupling effects between the individual cores, the range of multi-core fiber technology is degraded, but is still in the range of kilometers~\cite[p.~355]{SpaceDivision}. Mechanical limitations such as the minimal bend radius and the maximal force loads are stricter than for single-core fibers.

\subsection{Network Devices}

This section explains the network devices in an optical network that are relevant in this work. The components include devices that can generate, forward (route), and receive signals.

\subsubsection{Transceiver}
\label{sec:txrx}

Transceivers combine transmitter and receiver in one device. They provide a bidirectional interface between the network and a computer module. Their task is to generate the optical signals and convert them back into electronic information.

Besides the form factor and the connectors, especially the optical and electrical properties are relevant for the selection. The latter are irrelevant for this work and are, therefore, omitted.
\subsubsection{Transmitter}

The wave characteristics of the transmission are determined by the transmitter. Important key points are the wavelength, spectral width, and the transmission power. Further properties of the transmitter are wave modes, reflections, etc.~\cite[p.~4]{ds:transceiverDS}.

\subsubsection{Receiver}
For an optical signal to be received, the receiver has to be tuned to the appropriate wavelength. In addition, the polarization and power of the signal have to be compatible.
Light-sensitive semiconductors, such as photo-diodes, convert optical signals into electronic signals, which can be measured for data extraction. The received power has to be within the allowed range of the detector. If the power is too low, it is no not possible to distinguish between signal and thermal noise, i.e. the signal-to-noise ratio is too low \cite{FoaLoss}. If the received power is too large, the detector is overloaded. Modulations of the luminous flux cannot be detected. Overloading can cause permanent damage to the detector and should, therefore, be avoided. 

\subsubsection{Connectors}

Connectors attach optical cables to devices. Diverse connector types have been developed for a wide range of cable types and transmission technologies.
The transmission in the connector is lossy due to fiber interruption. Lenses, polishing of the ends, and special geometries are used to reduce the attenuation.

\subsection{Switches}
\label{sec:basics:opticalSwitches}

Switches form the heart of a network. Switches control signal routes between nodes. In classic copper networks, this routing is based on identifiers in the data packets. In contrast, routing in optical networks can also be based on the physical signal characteristics. This can be, for example, the wavelength in Wave-Division Multiplexing (WDM) or the core used in multi-core fiber transmission. 

\subsubsection{Opaque Switches}
Opaque switches convert the incoming optical signals to electrical signals for routing.
After signal conversion, the signals are routed through electrical conductors to a transmitter located at the output. The transmitter converts the signal back into an optical signal.
The advantages of opaque switches are digital signal processing and a flexibility of the routing. Incoming attenuated signals are amplified. In addition, electrical signals can be controlled without mechanical components, which simplifies routing. Dynamic route changes are easy to establish. 
However, a power supply is mandatory and the transmission speed in the optical network is limited by the electrical components. 
Opaque switches are easy to configure and cheaper than so-called translucent switches~\cite[p.~5]{Walker03transparentoptical}.

\subsubsection{Translucent Switches}

In translucent switches no conversion of the optical signal takes place. An incoming signal is routed through fixed optical fibers or miniature mirrors, depending on the technology. 
Advantages over opaque switches result from the fact that no conversion is required. Translucent switches are independent of the data rates \cite[p.~1]{ds:MemsAgiltronDs}. The signal transmission is delay-free over the entire wavelength range of the fibers used~\cite{ds:OXCSercalo}.

A disadvantage is the lack of signal conditioning. Signals are not amplified, which requires care in signal attenuation during path planning in network design.

Routing in translucent switches based on Micro-Electro-Mechanical Systems (MEMS) is accomplished via small, electromechanically adjustable micromirrors \cite[p.~13f.]{Walker03transparentoptical}. Dynamic routing can be established, but electrical power is required for the actuation.

An alternative are translucent switches with fixed optical fibers. In this case, the routing is directly specified by the physical connection from the input to the output, which limits this technology to static routing. Since this type does not contain any electrical or moving mechanical components, no power supply is required.

\subsection{Attenuation}
\label{sec:attenuation}
For a successful optical signal transmission, the attenuation has to be considered comprehensively. The possible transmission power of transceivers is limited as well as the input power range of receivers.

\subsubsection{Reasons for Attenuation}

An optical signal is attenuated by various disturbances and losses along the signal path. Physical effects such as scattering and absorption cause a majority of the attenuation within the fiber \cite{FoaOpticalFiber}.

All disturbances and impurities in a fiber lead to additional losses \cite{FoaLoss}. These losses are proportional to the signal power. Therefore, when considering attenuation losses, they are calculated in decibels. Therefore, attenuation losses are quantified in decibels and absolute quantities are specified in \si{\dBm} (decibel milliwatts).
All calculations in this work are carried out in decibels, which leads to an additive expression compatible to MILP. 
Additional attenuation is caused by connectors and splices. If the fiber ends are not exactly aligned, portions of the signal will get lost. This loss is called insertion loss \cite{FoaLoss}. In addition, reflections occur at the fiber ends, which also attenuates the signal.

\subsubsection{Design Calculation}

\begin{figure}[htbp]
	\centering
	
	\includegraphics[width=\linewidth]{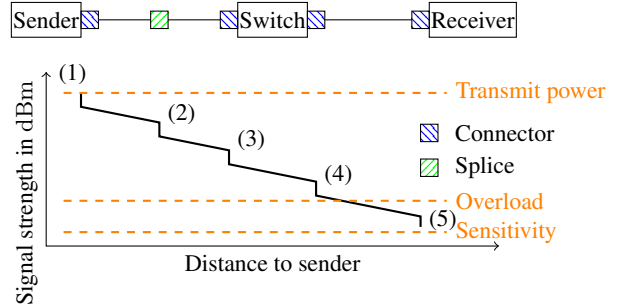}
	
	\caption{Exemplary signal strength curve (cf.~\cite{FoaLoss}). The switch is translucent in this example.}
	\label{fig:attenuationCalc}
\end{figure}

For the design purposes, all the attenuation that occur is accumulated and compared with the transmit and receive characteristics of the transceivers.
By means of fig. \ref{fig:attenuationCalc} the calculation is explained. In comparison to the presentation of the calculation in \cite{FoaLoss}, the connector losses at the transmitter and receiver are explicitly presented, since these are later assigned to the cables.
Starting from the transmitter, the signal is sent out with the transmit power (1). Within the connector, first losses occur in the form of an insertion loss and losses due to reflection. Along the cable, the transmit power reduces linearly with cable length due to dispersion and absorption effects. At the splice (2), fiber interference causes signal attenuation.
The transparent switch used in this example has an intrinsic attenuation, which is evident from the signal attenuation between input (3) and output (4). The switch is connected via connectors, which cause the abrupt signal degradation at points (3) and (4). The connector (5) connects the cable to the receiver. Since the signal is received after the connector, the connector attenuation also has to be considered.
The signal power at the receiver (5) has to be within the dynamic range, which is determined by the sensitivity and the overload limit. The signal in this example is receivable.
If (5) is below the sensitivity, thermal noise hides the signal. If it is above the overload limit, then the detector cannot process the signal.

%% file: Figures/problem.pdf_tex
\begingroup%
  \makeatletter%
  \providecommand\color[2][]{%
    \errmessage{(Inkscape) Color is used for the text in Inkscape, but the package 'color.sty' is not loaded}%
    \renewcommand\color[2][]{}%
  }%
  \providecommand\transparent[1]{%
    \errmessage{(Inkscape) Transparency is used (non-zero) for the text in Inkscape, but the package 'transparent.sty' is not loaded}%
    \renewcommand\transparent[1]{}%
  }%
  \providecommand\rotatebox[2]{#2}%
  \newcommand*\fsize{\dimexpr\f@size pt\relax}%
  \newcommand*\lineheight[1]{\fontsize{\fsize}{#1\fsize}\selectfont}%
  \ifx\svgwidth\undefined%
    \setlength{\unitlength}{382.31228085bp}%
    \ifx\svgscale\undefined%
      \relax%
    \else%
      \setlength{\unitlength}{\unitlength * \real{\svgscale}}%
    \fi%
  \else%
    \setlength{\unitlength}{\svgwidth}%
  \fi%
  \global\let\svgwidth\undefined%
  \global\let\svgscale\undefined%
  \makeatother%
  \begin{picture}(1,0.7664365)%
    \lineheight{1}%
    \setlength\tabcolsep{0pt}%
    \put(0,0){\includegraphics[width=\unitlength,page=1]{problem.pdf}}%
    \put(0.14189608,0.68922995){\color[rgb]{0,0,0}\makebox(0,0)[lt]{\lineheight{1.25}\smash{\begin{tabular}[t]{l}Signal A\end{tabular}}}}%
    \put(0.14252054,0.65793316){\color[rgb]{0,0,0}\makebox(0,0)[lt]{\lineheight{1.25}\smash{\begin{tabular}[t]{l}Signal B\end{tabular}}}}%
    \put(0,0){\includegraphics[width=\unitlength,page=2]{problem.pdf}}%
    \put(0.14227519,0.62895596){\color[rgb]{0,0,0}\makebox(0,0)[lt]{\lineheight{1.25}\smash{\begin{tabular}[t]{l}Signal C\end{tabular}}}}%
    \put(0,0){\includegraphics[width=\unitlength,page=3]{problem.pdf}}%
    \put(0.48938809,0.69795518){\color[rgb]{0,0,0}\makebox(0,0)[lt]{\lineheight{1.25}\smash{\begin{tabular}[t]{l}Signal A\end{tabular}}}}%
    \put(0.4900126,0.6666584){\color[rgb]{0,0,0}\makebox(0,0)[lt]{\lineheight{1.25}\smash{\begin{tabular}[t]{l}Signal B\end{tabular}}}}%
    \put(0,0){\includegraphics[width=\unitlength,page=4]{problem.pdf}}%
    \put(0.48976721,0.6376812){\color[rgb]{0,0,0}\makebox(0,0)[lt]{\lineheight{1.25}\smash{\begin{tabular}[t]{l}Signal C\end{tabular}}}}%
    \put(0,0){\includegraphics[width=\unitlength,page=5]{problem.pdf}}%
    \put(0.62737231,0.23420021){\color[rgb]{0,0,0}\makebox(0,0)[lt]{\lineheight{1.25}\smash{\begin{tabular}[t]{l}Cables ?\end{tabular}}}}%
    \put(0,0){\includegraphics[width=\unitlength,page=6]{problem.pdf}}%
    \put(0.6271539,0.58723339){\color[rgb]{0,0,0}\makebox(0,0)[lt]{\lineheight{1.25}\smash{\begin{tabular}[t]{l}Switches ?\end{tabular}}}}%
    \put(0.62693549,0.05002541){\color[rgb]{0,0,0}\makebox(0,0)[lt]{\lineheight{1.25}\smash{\begin{tabular}[t]{l}- Attenuation?\\- Direction?\end{tabular}}}}%
    \put(0.62693549,0.39858099){\color[rgb]{0,0,0}\makebox(0,0)[lt]{\lineheight{1.25}\smash{\begin{tabular}[t]{l}- Number of ports?\\- Space restriction?\\\end{tabular}}}}%
    \put(0.84036277,0.5689406){\color[rgb]{0,0,0}\makebox(0,0)[lt]{\lineheight{1.25}\smash{\begin{tabular}[t]{l}\\Opaque \\or \\translucent?\end{tabular}}}}%
    \put(0,0){\includegraphics[width=\unitlength,page=7]{problem.pdf}}%
    \put(0.62869419,0.57062516){\color[rgb]{0,0,0}\makebox(0,0)[lt]{\lineheight{1.25}\smash{\begin{tabular}[t]{l}\\Type 1\\\\Type 2\\...\end{tabular}}}}%
    \put(0.62869419,0.21898602){\color[rgb]{0,0,0}\makebox(0,0)[lt]{\lineheight{1.25}\smash{\begin{tabular}[t]{l}\\Type 1\\\\Type 2\\...\end{tabular}}}}%
    \put(0.87737864,0.18926097){\color[rgb]{0,0,0}\makebox(0,0)[lt]{\lineheight{1.25}\smash{\begin{tabular}[t]{l}2 \\or\\4 cores\\or ...? \end{tabular}}}}%
    \put(0.28950711,0.19891418){\color[rgb]{0,0,0}\makebox(0,0)[lt]{\lineheight{1.25}\smash{\begin{tabular}[t]{l}Maximum \\topology\end{tabular}}}}%
    \put(0.72956025,0.73203119){\color[rgb]{0,0,0}\makebox(0,0)[lt]{\lineheight{1.25}\smash{\begin{tabular}[t]{l}End system\\with communication \\needs\end{tabular}}}}%
    \put(0,0){\includegraphics[width=\unitlength,page=8]{problem.pdf}}%
    \put(0.6271539,0.30239012){\color[rgb]{0,0,0}\makebox(0,0)[lt]{\lineheight{1.25}\smash{\begin{tabular}[t]{l}Signal routing ?\end{tabular}}}}%
    \put(0,0){\includegraphics[width=\unitlength,page=9]{problem.pdf}}%
  \end{picture}%
\endgroup%

%% file: Figures/2-3_multi_core.pdf_tex
\begingroup%
  \makeatletter%
  \providecommand\color[2][]{%
    \errmessage{(Inkscape) Color is used for the text in Inkscape, but the package 'color.sty' is not loaded}%
    \renewcommand\color[2][]{}%
  }%
  \providecommand\transparent[1]{%
    \errmessage{(Inkscape) Transparency is used (non-zero) for the text in Inkscape, but the package 'transparent.sty' is not loaded}%
    \renewcommand\transparent[1]{}%
  }%
  \providecommand\rotatebox[2]{#2}%
  \newcommand*\fsize{\dimexpr\f@size pt\relax}%
  \newcommand*\lineheight[1]{\fontsize{\fsize}{#1\fsize}\selectfont}%
  \ifx\svgwidth\undefined%
    \setlength{\unitlength}{400bp}%
    \ifx\svgscale\undefined%
      \relax%
    \else%
      \setlength{\unitlength}{\unitlength * \real{\svgscale}}%
    \fi%
  \else%
    \setlength{\unitlength}{\svgwidth}%
  \fi%
  \global\let\svgwidth\undefined%
  \global\let\svgscale\undefined%
  \makeatother%
  \begin{picture}(1,0.575)%
    \lineheight{1}%
    \setlength\tabcolsep{0pt}%
    \put(0,0){\includegraphics[width=\unitlength,page=1]{2-3_multi_core.pdf}}%
    \put(0.14861694,0.3435413){\color[rgb]{0,0,0}\makebox(0,0)[t]{\lineheight{1.25}\smash{\begin{tabular}[t]{c}Multi-core fiber\end{tabular}}}}%
    \put(0.84861691,0.3435413){\color[rgb]{0,0,0}\makebox(0,0)[t]{\lineheight{1.25}\smash{\begin{tabular}[t]{c}Multi-core fiber\end{tabular}}}}%
    \put(0.37893554,0.41911252){\color[rgb]{0,0,0}\makebox(0,0)[t]{\lineheight{1.25}\smash{\begin{tabular}[t]{c}Switch\end{tabular}}}}%
    \put(0,0){\includegraphics[width=\unitlength,page=2]{2-3_multi_core.pdf}}%
    \put(0.22856445,0.11716708){\color[rgb]{0,0,0}\makebox(0,0)[lt]{\lineheight{1.25}\smash{\begin{tabular}[t]{l}Cable/fiber \hspace{0.3cm} Signal path\end{tabular}}}}%
    \put(0.22849731,0.06284917){\color[rgb]{0,0,0}\makebox(0,0)[lt]{\lineheight{1.25}\smash{\begin{tabular}[t]{l}Cores\end{tabular}}}}%
    \put(0,0){\includegraphics[width=\unitlength,page=3]{2-3_multi_core.pdf}}%
  \end{picture}%
\endgroup%

%% file: related_work.tex
\section{Related Work}
\label{sec:related_work}

Optimization is a common issue for (optical) networks. Typical subjects of optimization are routing \cite{sheikh2012, carta2012, INPROC-2018-46, Smirnov:2017:OMR:3061639.3062298, Singh2017Routingalgorithms, EBRAHIMZADEH2013354, 10.1007/978-981-10-3325-4_6, Ayoub2020}, scheduling \cite{INPROC-2016-32, INPROC-2018-46, Smirnov:2017:OMR:3061639.3062298}, robustification \cite{Ayoub2020, Ergenc2021}, and frequency sharing \cite{EBRAHIMZADEH2013354, Zorello2020}. 
The mentioned works are exemplary and the list is not complete. Some of the works address multiple issues at once. MILP or ILP are common methods to solve the problems. Whereas the existing solutions for routing match our required routing from defined senders to defined receives with respect to resource constraints, the underlying network topology needs to be predefined. That is not the case for our problem. Topology optimization is a less frequent research topic, but it has been addressed for electrical \cite{acevedo2012,OptimierungBordnetzarchitektur2000,Li2018,gupta2006} and optical networks \cite{IncrementalOptical2011, Rahman2016, Filippini2010, Agata2012, Sousa2016, Youssef2010, Kokangul2011}. The approaches for electrical networks omit optical properties such as attenuation and multi-core cables. The approaches for optical networks are very domain-specific. They deal with special applications (e.g. telecommunication, dynamic routing) or certain optical network technologies (e.g. WDM), that do not match our setup, i.e. static routing in a restricted space while choosing optimal cables, switches, and topology. In no existing approach the placement and choice of components is subject to optimization in parallel. The approach most similar to our needs is \cite{AnnighoeferArticle}. It suggests a Binary Program representation for a topology optimization of Ethernet-like networks. The approach considers switch types, port, and resource restrictions as well as signal routing and signal segregation constraints. We extend that approach to optical networks by adding multi-core cable types and attenuation.

%% file: theory.tex
\section{A Mixed-integer Linear Programming Representation for Topology Optimization}
\label{sec:theory}

The topology optimization problem is expressed as a MILP, whose objective function and constraints are linear and contain both binary, integer, and continuous variables: 

\begin{align}
	&\text{maximize} \nonumber\\
	&\sum_{j \in B} c_{j} x_{j} + \sum_{j \in I} c_{j} x_{j} + \sum_{j \in C} c_{j} x_{j} \label{eq:milp:cost}\\
	&\text{subject to} \nonumber\\
	&\sum_{j \in B} a_{ij} x_{j} + \sum_{j \in I} a_{ij} x_{j} + \sum_{j \in C} a_{ij} x_{j} %
		\begin{Bmatrix}
			\leq\\ 
			= \\
			\geq
		\end{Bmatrix}%
		b_i, \forall i \in M \label{eq:milp:constr}\\	
	 &l_j \leq x_j \leq u_j  \qquad \forall j \in N = B \cup I \cup C \label{eq:milp:3}\\
	 &x_j \in \binary, \forall j \in B \label{eq:milp:4}\\
	 &x_j \in \integers, \forall j \in I \label{eq:milp:5}\\
	 &x_j \in \reals, \forall j \in C  \label{eq:milp:6}
\end{align}

The solution $x$ is a set of variables $x_j$,~$j \in N$ that satisfies the conditions (\ref{eq:milp:constr})-(\ref{eq:milp:6}). The aim is to maximize an objective function (\ref{eq:milp:cost}). This is a weighted sum with cost factors $c_j$,~$j \in N$ \cite[p.~2]{Hoffman2013}. 
The equalities and inequalities (\ref{eq:milp:constr}) are constraints \cite[p.~84]{Kallrath2013}. Each comprises coefficients $a_{ij}$. The indices associated with the constraints are included in the set $M$.
The sets $B$, $I$, and $C$ are the index sets of binary, integer, and continuous variables \cite[p.~2]{Hoffman2013}. 
Formally, the variable data type is specified by the equations (\ref{eq:milp:4}), (\ref{eq:milp:5}), and (\ref{eq:milp:6}). (\ref{eq:milp:3}) allows to set additional upper ($u_j$) and lower ($l_j$) bounds for $x_j$, $j \in N$.
The cost function (\ref{eq:milp:cost}) is maximized. Problems that require the objective function to be minimized are addressed by inverting the const function:
\begin{equation}
\label{eq:milp:minimal}
\min \left \{ f(x) \right \} \equiv \max \left \{ -f(x) \right \}
\end{equation}

Essential for topology optimization is the encoding of the expected topology in the solution vector 

\begin{align}
&\var{}{} = (\var{D_k}{\set T_l},\ldots \qquad \in \binary \label{eq:vars:device_type} \\
&, \var{F_j}{\set T_t},\ldots \qquad \in \binary \label{eq:vars:cable_type} \\ 
&, \var*{F_j}{useAB}, \var*{F_j}{useBA} \qquad \in \integers \label{eq:vars:fiber_count} \\
&, \var*{F_j}{allowAB}, \var*{F_j}{allowBA}, \ldots  \qquad \in \binary \label{eq:vars:direction} \\
&, \var{S_i}{F_j , \text{AB}}, \var{S_i}{F_j , \text{BA}}, \ldots \in \binary \qquad \label{eq:vars:signal_to_cable} \\
&, \var{S_i}{D_k , \text{doesRx}}, \var{S_i}{D_k , \text{doesTx}}, \var{S_i}{D_k , \text{Rx}}, \var{S_i}{D_k , \text{transm.}} \in \binary \label{eq:vars:signal_to_device} \\
&, \var{S_i}{D_k , \text{TxAvail}}, \var{S_i}{D_k , \text{Tx}}, \var{S_i}{F_j , \text{powerAB}},  \var{S_i}{F_j , \text{powerAB}} \in \reals \label{eq:vars:attenuation}\\
&, \var{S_i}{D_k , \text{opaqueRx}} \qquad \in \binary ). \label{eq:vars:opaque}
\end{align} 
 
\var{}{} is based on the fact that the maximum topology is calculated a priori from the given installation space, i.e. the maximum number of switch devices and links is known. 
Solving the MILP extracts the optimum from the maximum topology.

Variables (\ref{eq:vars:device_type}) and (\ref{eq:vars:cable_type}) encode the existence of all possible devices and cables of their chosen type. 
The set $\set D$ denotes all possible devices (switches) in the network. The element $D_k \in \set D$ is the $i$th device. The total number of devices possible is denoted by $\abs{\set D}$.
Further, $\set F$ represents the set of all potential cables. This includes those cables automatically generated to form a fully connected network. A single cable of this set is referred to as $F_j \in \set F$. $\abs{\set F}$ is the total number of possible cables.
The notation $F_j = \left( D_A, D_B \right)$ expresses which two devices $D_A$ and $D_B$ are connected by cable $F_j$, with $D_A \neq D_B$.
Two cables $F_m$ and $F_n$ linking the same two devices $D_A$ and $D_B$ are legal. Furthermore, the direction is considered, i.e. $\left( D_A, D_B \right) \neq \left( D_B, D_A \right)$. 

Signals are represented by the set $\set S$, the signal count is $\abs{\set S}$. Source and target $D_S$ and $D_T$ of a signal $S_i \in \set S$ are indicated by the notation $S_i = \left( D_S, D_T \right)$. Multiple signals are allowed between the same source and target device, but signal aggregation is recommended as each signal requires its own core assignment. Bandwidth limitations have to be considered in signal aggregation.

\subsection{Device, Cable types, and Properties}
\label{sec:algo:typing}
Devices and cables are assigned with property vectors. A vector contains all the quantities of the type. Class $\set E$ is a representative of device set $\set D$ or cable set $\set F$.

\subsubsection{Type declaration}
\label{sec:algo:typing:declare}

A type $t$ belonging to the element class $\set E$ is represented by the constant property vector	\begin{equation}
	\vec{\tau}{\set E}{t} =
	\transpose{\vec*{
			\place{\tau}{\set E}{1_t} & \cdots &
			\place{\tau}{\set E}{i_t} & \cdots &
			\place{\tau}{\set E}{{\abs{\vec{\tau}{\set E}}}_t}
	}},
	\vec{\tau}{\set E}{t} \in \reals^{\abs{\vec{\tau}{\set E}}}.
	\end{equation}
$\abs{\vec{\tau}{\set E}}$ denotes the number of properties that are identical for all representatives of the element class.
In general, each property is a real value. Sometimes a restriction to discrete values is useful.
Across all types, $\vec{\hat{\tau}}{\set E}{i}$ gives the maximum value of a property $\vec{\tau}{\set E}{i}$. $\vec{\check\tau}{\set E}{i}$ is the minimum value.

All representatives of the element class can be combined into a constant matrix $\matrix{T}{\set E}$. The number of types of the element class $\set E$ is given by $\abs{\matrix{T}{\set E}}$.

\begin{equation}
\matrix{T}{\set E} = \matrix*{%
	\vec{\tau}{\set E}{1} & \cdots &
	\vec{\tau}{\set E}{\abs{\matrix{T}{\set E}}}%
},%
\matrix{T}{\set E} \in \reals^{\abs{ \vec{\tau}{\set E} } \times \abs{\matrix{T}{\set E}}}
\end{equation}

\subsubsection{Type Assignment}
\label{sec:algo:typing:assign}

For the assignment of a type to an element $e \in \set E$, this has a type vector $\var{e}{\set T}$ which encodes the actual type. A one in the vector entry $\var{e}{\set{T}_{t}}$ means the element $e$ is of type $t$. A zero corresponds to saying that the element is not of the associated type $t$. Other values are not allowed in

\begin{equation}
\var{e}{\set T} = \transpose{\vec*{
		\var{e}{\set{T}_{1}} \cdots
		\var{e}{\set{T}_{t}} \cdots
		\var{e}{\set{T}_{\abs{\matrix{T}{\set E} }}}
}},
\var{e}{\set T} \in \binary^{\abs{ \matrix{T}{\set E} }}.
\end{equation}

For unambiguous type assignment, the choice of the type vector $\var{e}{\set T}$ is restricted to have at most a single entry equal to one. A limiting SOS constraint \cite[147~ff]{Kallrath2013} is used. A suitable choice of the (in)equalities determines whether the element $e$ has to exist by definition or can be removed by optimization.

If an element has to exist this means that a type is assigned to it. The SOS equation \ref{eq:type:sosEq} specifies the type assignment obligation.%
\begin{equation}
\sum_{t = 1}^{\abs{\matrix{T}{\set E}}} \var{e}{\set{T}_{t}} = 1
\label{eq:type:sosEq}
\end{equation}

The elements that exist according to definition are compared with the optional elements. An element to which no type is assigned has all assignment entries set to zero. 
For this the following SOS equation ensures a valid assignment:
\begin{equation}
\sum_{t = 1}^{\abs{\matrix{T}{\set E}}} \var{e}{\set{T}_{t}} \leq 1
\label{eq:type:sosUeq}
\end{equation}

In addition to defining whether an element must exist in the final topology, the ability to restrict the allowed types is required. This is done by specifying forbidden mappings $t$:
\begin{equation}
\var{e}{\set{T}_{t}} = 0
\end{equation}

In the following it is necessary to know whether an element exists in the network or not. This is given by sum in the two conditions (\ref{eq:type:sosEq}) and (\ref{eq:type:sosUeq}). This notation is used as an abbreviation in the following.
\begin{equation}
	\sumNorm{\var{e}{\set T}} := \sum_{t = 1}^{\abs{\matrix{T}{\set E}}} \var{e}{\set{T}_{t}}
\end{equation}

\subsubsection{Element properties}
\label{sec:algo:typing:property}

From the type matrix $\matrix{T}{\set E}$ and the type vector \var{e}{\set T} of the element $e$, the current properties of the element are determined with (\ref{eq:type:prop}). These are stored in an element property vector $\property{e}{}$.
\begin{equation}
\property{e}{} =
\matrix{T}{\set E} \cdot \var{e}{\set T} =
\transpose{\vec*{
		\property{e}{1} \cdots 
		\property{e}{i} \cdots
		\property{e}{\abs{\vec{\tau}{\set E}}}
}},
\property{e}{} \in \reals^{\abs{\vec{\tau}{\set E}}}
\label{eq:type:prop}
\end{equation}
For a property $i$ the equation is:
\begin{equation}
\property{e}{i} =
\vec{\tau}{\set E}{i_1} \cdot \var{e}{\set{T}_{1}} +
\cdots +
\vec{\tau}{\set E}{i_{\abs{\matrix{T}{\set E}}}} \cdot \var{e}{\set{T}_{\abs{\matrix{T}{\set E}}}}
\end{equation}

If the element does not exist, i.e. $\var{e}{\set T} = 0$, then all element properties $\property{e}{i} = 0$. 
\subsubsection{Device Properties}
\label{sec:algo:typing:actual:devices}

The device properties listed in table \ref{tbl:properties:devices} are essential to consider the requirements on attenuation and routing constraints.

\begin{table}[h!tbp]
	\centering

	\renewcommand{\arraystretch}{1.5}
	\begin{tabular}{@{}llll@{}}
		\toprule
		\textbf{Property} & \textbf{Symbol} & \textbf{Type} & \textbf{Unit}\\	
		\midrule
		
		Port count & $\property*{D_k}{ports}$ & $\integers ^ + $ & --- \\
		Internal signal attenuation & $\property{D_k}{\Delta}$ & \reals & \si{\decibel} \\
		Minimum received power & $\property{D_k}{\underlineText{Rx}}$ & \reals & \si{\dBm} \\
		Maximum received power & $\property{D_k}{\overlineText{Rx}}$ & \reals & \si{\dBm} \\
		Minimum transmit power & $\property{D_k}{\underlineText{Tx}}$ & \reals & \si{\dBm} \\
		Maximum transmission power & $\property{D_k}{\overlineText{Tx}}$ & \reals & \si{\dBm} \\
		Translucent transmission & $\property*{D_k}{trans}$ & \binary & --- \\
		Cost value & $\property*{D_k}{cost}$ & $\reals ^ +$ & --- \\
		
		\midrule
		Type vector & $\var{D_k}{\set T} $ & $\binary^{\abs{ \matrix{T}{\set D} }}$ & --- \\
		\bottomrule
	\end{tabular}
	
	\caption{Overview of the device properties of a device $D_k \in \set D$.}
	\label{tbl:properties:devices}
\end{table}

The port count $\property*{D_k}{ports}$ limits the number of available cable ports. It is the main resource constraint. Whereas, the five properties $\property{D_k}{\Delta}$, $\property{D_k}{\underlineText{Rx}}$, $\property{D_k}{\overlineText{Rx}}$, $\property{D_k}{\underlineText{Tx}}$ and $\property{D_k}{\overlineText{Tx}}$ are used to account for attenuation. For these properties, the following conditions are imposed depending on the translucent property $\property*{D_k}{trans}$:%
\begin{align}
	\property*{D_k}{trans} &= 0 &\Rightarrow&&%
	\property{D_k}{\Delta} &= 0
	\label{eq:type:device:opaque}\\
	\property*{D_k}{trans} &= 1 &\Rightarrow&&%
	\property{D_k}{\underlineText{Rx}} =
	\property{D_k}{\overlineText{Rx}} =
	\property{D_k}{\underlineText{Tx}} =
	\property{D_k}{\overlineText{Tx}} &= 0
	\label{eq:type:device:trans}
\end{align}

Devices $D_k$ for which $\property*{D_k}{trans} = 1$ are translucent
 devices. If $\property*{D_k}{trans} = 0$, the device is opaque.

Further, it is required that the maximum and minimum values of the transmit and receive power are chosen as such:
\begin{align}
	\property{D_k}{\underlineText{Rx}} \leq \property{D_k}{\overlineText{Rx}}
		\label{eq:type:device:consistency1}\\
	\property{D_k}{\underlineText{Tx}} \leq \property{D_k}{\overlineText{Tx}}
		\label{eq:type:device:consistency2}
\end{align}

The conditions (\ref{eq:type:device:opaque}) to (\ref{eq:type:device:consistency2}) are applied to the property values of the device types. The conditions can trivially be satisfied by mindful device type definition.

The last property $\property*{D_k}{cost}$ is solely for the cost function. 

\subsubsection{Cable Properties}
\label{sec:algo:typing:actual:cables}

Variables related to cable type and thus cable properties are listed in table \ref{tbl:properties:cables}. 

The core count $\property*{F_j}{cores}$ and the signal attenuation $\property{F_j}{\Delta}$ are the most important properties of an optical cable.
The cost value $\property*{F_j}{cost}$ is in use exclusively in the cost function. The cable type vector $\var{F_j}{\set T}$ is included for completeness.
\begin{table}[htbp]
	\centering

	\renewcommand{\arraystretch}{1.5}
	\begin{tabular}{@{}llll@{}}
		\toprule
		\textbf{Property} & \textbf{Symbol} & \textbf{Type} & \textbf{Unit}\	\\
		\midrule
		
		Number of cores & $\property*{F_j}{cores}$ & $\integers ^ +$ & --- \\
		Attenuation & $\property{F_j}{\Delta}$ & \reals & \si{\decibel} \\
		Cost value & $\property*{F_j}{cost}$ & $\reals ^ +$ & --- \\
		
		\midrule
		\multicolumn{4}{@{}l}{\textbf{transmission constraints}} \\
		Unidirectional & $\property*{F_j}{uni}$ & \binary & --- \\
		Direction $A \rightarrow B$ allowed & $\property*{F_j}{AB}$ & \binary & --- \\
		Direction $B \rightarrow A$ allowed & $\property*{F_j}{BA}$ & \binary & --- \\

		\midrule
		Type vector & $\var{F_j}{\set T} $ & $\binary^{\abs{ \matrix{T}{\set F} }}$ & --- \\
		\bottomrule
	\end{tabular}
	
	\caption{Overview of cable properties of a cable $F_j \in \set F$.}
	\label{tbl:properties:cables}
\end{table}

The transmission direction of a cable is controlled by the properties $\property*{F_j}{uni}$, $\property*{F_j}{AB}$ and $\property*{F_j}{BA}$.

The index $AB$ indicates the direction of transmission from devices $D_A$ to $D_B$. The index $BA$ denotes the opposite direction. This convention is used below for other direction-related variables.
\begin{table}[htbp]
	\centering
	
	\begin{tabular}{@{}cccl@{}}
		\toprule
		$\property*{F_j}{uni}$ & $\property*{F_j}{AB}$ & $\property*{F_j}{BA}$ & \textbf{Explanation} \\
		\midrule
		
		\num{0} & \num{1} & \num{1} & bidirectional \\
		\num{1} & \num{1} & \num{1} & Unidirectional, no direction specified\\
		\num{1} & \num{1} & \num{0} & Unidirectional, direction $A \rightarrow B$ \\
		\num{1} & \num{0} & \num{1} & Unidirectional, direction $B \rightarrow A$ \\
		
		\midrule
		
		$\bullet$ & \num{0} & \num{0} & Not allowed, no direction allowed \\
		\num{0} & \num{0} & \num{1} & Not allowed, inconsistent\\
		\num{0} & \num{1} & \num{0} & Not allowed, inconsistent\\
		
		\bottomrule
	\end{tabular}
	
	\caption{Allowed and disallowed values for the directionality properties of a cable~$F_j$.}
	\label{tbl:properties:cables:direction}
\end{table}

The property $\property*{F_j}{uni}$ describes whether the cable $F_j$ can transmit signals bidirectionally (value zero) or unidirectionally (value one). The two properties $\property*{F_j}{AB}$ and $\property*{F_j}{BA}$ define the allowed directions. If the value is one, a transmission in the associated direction is allowed. Consistency between unidirectionality and allowed directions has to be ensured. Table \ref{tbl:properties:cables:direction} lists all valid values.

Although a cable may only transmit unidirectionally, the direction is not restricted ($\property*{F_j}{uni} = \property*{F_j}{AB} = \property*{F_j}{BA} = 1$). This can be used to model unidirectional cables where the transmission direction is not specified in advance.

\subsection{Port Count Limits}

The number of ports of a device $D_k \in \set D$ limits the number of cables. No more cables can be connected than there are ports.

Cables $F_j$ whose definition contains the device $D_k$ as start or end points are relevant for the constraint. This is true for the subset~$\left\{ \set F \mid F_j = \left(D_k, \ast \right) \lor F_j = \left(\ast, D_k \right) \right\} \subseteq \set F$.
Since an existing cable have to be connected to the device, the sum of all existing cables is the number of connected cables. A sum norm of the type vector equal to one describes the existence of a cable $F_j$. 
The inequality \ref{eq:resources:ports} relates the number of connected cables to the number of ports of the device $D_k$.%
\begin{equation}
\sum_{F_j \in
	\left\{ \set F \mid F_j = \left(D_k, \ast \right) \lor F_j = \left(\ast, D_k \right) \right\}
}%
\sumNorm{\var{F_j}{\set T}}%
\leq \property*{D_k}{ports}%
\label{eq:resources:ports}
\end{equation}

This constraint also prohibits the connection of cables of devices that do not exist. In this case $\property*{D_k}{ports} = 0$, which prohibits all potential cables. Conversely, this requires that devices connected by a fixed cable always have to have ports and thus have to be existing device instances.

\subsection{Directionality and Core Count Limits}
\label{sec:algo:resources:directionLimit}

Another resource constraint is the directionality of the cables as well as the maximum number of transmissions possible. The variables listed in table \ref{tbl:resources:cables:var} are necessary for these constraints.
\begin{table}[htbp]
\small
	\caption{List of variables of each cable $F_j \in \set F$. These are required to set up the constraints}
	\centering

	\renewcommand{\arraystretch}{1.5}
	\begin{tabular}{@{}lll@{}}
		\toprule
		\textbf{Designation} & \textbf{Symbol} & \textbf{Type}\\	
		\midrule
		
		Transfer count $A \rightarrow B$ & $\var*{F_j}{useAB}$ & $\integers ^ +$ \\
		Transfer count $B \rightarrow A$ & $\var*{F_j}{useBA}$ & $\integers ^ +$ \\
		
		Transfer allowance $A \rightarrow B$ & $\var*{F_j}{allowAB}$ & \binary \\
		Transfer permit $B \rightarrow A$ & $\var*{F_j}{allowBA}$ & \binary \\
		
		\bottomrule
	\end{tabular}

	\label{tbl:resources:cables:var}
\end{table}
The total number of transmissions is limited by the core count $\property*{F_j}{cores}$ of a cable $F_j \in \set F$. Each core may only transmit a single signal. The number of transmissions is counted for each cable by the variables $\var*{F_j}{useAB}$ and $\var*{F_j}{useBA}$. The sum of the count variables gives the total count, which is limited by the core count as follows:
\begin{equation}
\var*{F_j}{useAB} + \var*{F_j}{useBA} \leq \property*{F_j}{cores} \ .
\end{equation}


Directionality is modeled by limiting the count variables depending on the directionality properties.
The two auxiliary binary variables $\var*{F_j}{allowAB}$ and $\var*{F_j}{allowBA}$ are introduced to limit the direction of unidirectional cables.
The two inequalities \ref{eq:resources:dir:ab} and \ref{eq:resources:dir:ba} limit the direction depending on the type definition for unidirectional cables.
If a direction is forbidden for the selected type, the associated property is zero, which also forbids the corresponding auxiliary variable to be used. For bidirectional types it is required that both directions are allowed.
\begin{align}
	\var*{F_j}{allowAB} &\leq \property*{F_j}{AB}
		\label{eq:resources:dir:ab}\\
	\var*{F_j}{allowBA} &\leq \property*{F_j}{BA}
		\label{eq:resources:dir:ba}
\end{align}

Equation \ref{eq:resources:dirUni} controls direction depending on whether the cable is allowed to be used bidirectionally or unidirectionally. 

\begin{equation}
\var*{F_j}{allowAB} + \var*{F_j}{allowBA} =
	\num{2} \cdot \sumNorm{\var{F_j}{\set T}}
	- \property*{F_j}{uni}
\label{eq:resources:dirUni}
\end{equation}

To explain the equation, three cases are considered:
\begin{enumerate}
	\item
		No type is assigned to the cable, it is $\sumNorm{\var{F_j}{\set T}} = \property*{F_j}{uni} = 0$. No transfers are allowed at all. This is consistent with the assumption of a non-existent cable.
		
	\item
		The type associated with the cable allows bidirectional transmission. Consequently, $\sumNorm{\var{F_j}{\set T}} = 1$ and $\property*{F_j}{uni} = 0$ holds, i.e.
		\begin{equation*}
			\var*{F_j}{allowAB} + \var*{F_j}{allowBA} = \num{2},
		\end{equation*}
		so that both directions are allowed.		
	
	\item
		In this case, the cable is assigned a type with unidirectional transmission $\sumNorm{\var{F_j}{\set T}} = \property*{F_j}{uni} = 1$. Substituting (\ref{eq:resources:dirUni}) gives:
		\begin{equation*}
		\var*{F_j}{allowAB} + \var*{F_j}{allowBA} = \num{1}.
		\end{equation*}
		Unless there is a restriction on the direction via (\ref{eq:resources:dir:ab}) or (\ref{eq:resources:dir:ba}), the MILP solver is free to determine the direction of the unidirectional cable.
	
\end{enumerate}

The final limit for the count variables is set using the inequalities \ref{eq:resources:dir:limit1} and \ref{eq:resources:dir:limit2}. If the cable is forbidden, the count variable is forced to zero, i.e., no transfers. If transfers are allowed in that direction, the bound is raised to the largest allowed transfer count. This corresponds to the maximum core count across all cable types, described by the constant $\vec{\hat{\tau}}{\set F}{\text{cores}}$.%
\begin{align}
	\var*{F_j}{useAB} &\leq \var*{F_j}{allowAB} \cdot \vec{\hat{\tau}}{\set F}{\text{cores}}
		\label{eq:resources:dir:limit1}\\
	\var*{F_j}{useBA} &\leq \var*{F_j}{allowBA} \cdot \vec{\hat{\tau}}{\set F}{\text{cores}}
		\label{eq:resources:dir:limit2}
\end{align}

\subsection{Routing}
\label{sec:algo:routing}

A valid routing of a signal $S_i \in \set S$, $S_i = \left(D_S, D_T\right)$ consists of a continuous, i.e., uninterrupted, path from the source device $D_S$ to the target device $D_T$.

The variables listed in table \ref{tbl:routing:cables:var} encode the path of signal $S_i$. The path encodes the devices $D_k \in \set D$ that transmit or receive the signal as well as the allocations to directed cables $F_j \in \set F$. If the value is one, the signal is transmitted over the device or cable. If it is zero, the signal does not pass over it.

\begin{table}[htbp]
	\centering

	\scriptsize
	\begin{tabular}{@{}llll@{}}
		\toprule
		\textbf{Designation} & \textbf{Symbol} & \textbf{Type} & \textbf{Exists for}\\	
		\midrule
		
		$F_j$ in direction $A \rightarrow B$ & $\var{S_i}{F_j , \text{AB}}$ & \binary & $\forall S_i \in \set S$, $\forall F_j \in \set F$ \\
		$F_j$ in direction $B \rightarrow A$ & $\var{S_i}{F_j , \text{BA}}$ & \binary & $\forall S_i \in \set S$, $\forall F_j \in \set F$ \\
		
		$D_k$ receives signal & $\var{S_i}{D_k , \text{doesRx}}$ & \binary & $\forall S_i \in \set S$, $\forall D_k \in \set D$ \\
		$D_k$ transmits signal & $\var{S_i}{D_k , \text{doesTx}}$ & \binary & $\forall S_i \in \set S$, $\forall D_k \in \set D$ \\
						
		\bottomrule
	\end{tabular}
	
	\caption{Each signal $S_i \in \set S$ has the listed variables, which contain information for the routing constraints. Each listed variable is required multiple times; once for each element listed in the last column.}
	\label{tbl:routing:cables:var}
\end{table}
\subsubsection{Start and End of a Signal Path}
\label{sec:algo:routing:startEnd}

Translucent devices cannot be the source or destination of a signal. Therefore, the source and target devices $D_S$ and $D_T$ of a signal $S_i = \left(D_S, D_T\right)$ have to be opaque. This is represented by two constraints:

\begin{align}
	\property*{D_S}{trans} &= \num{0} \\
	\property*{D_T}{trans} &= \num{0}
\end{align}

For the source device $D_S$ it is required that the signal is emitted by it. In addition, it is specified that the source device does not receive the signal. On the other hand, it has to be specified that the target device $D_T$ receives the signal and does not transmit it:
\begin{align}
\var{S_i}{D_A , \text{doesTx}} &= \num{1}\\
\var{S_i}{D_A , \text{doesRx}} &= \num{0}\\
\nonumber\\
\var{S_i}{D_B , \text{doesTx}} &= \num{0}\\
\var{S_i}{D_B , \text{doesRx}} &= \num{1}
\end{align}

\subsubsection{Intermediate Devices}

Regarding other devices, a device $D_k \in \set D \setminus \left\{D_S, D_T\right\}$ can transmit the signal only if it was received before:
\begin{equation}
	\var{S_i}{D_k , \text{doesTx}} = \var{S_i}{D_k , \text{doesRx}}
\end{equation}

Either the device transmits and receives or does nothing. The coupling of devices is encoded with the variables $\var{S_i}{F_j , \text{AB}}$ and $\var{S_i}{F_j , \text{BA}}$ from table \ref{tbl:routing:cables:var}. They encode whether the signal $S_i$ is transmitted over the cable $F_j$ and thus the two devices are connected by that cable.

\subsubsection{Device-cable-coupling}

The coupling between the individual devices is expressed by the cable usage. Since cables have a direction, this must be considered when setting up the constraints.
According to table \ref{tbl:routing:cables:var} each signal $S_i$ for each cable $F_j$ has two directional variables that encode the transmission direction of the signal.
\begin{figure}[ht]
	\centering
	\def\svgwidth{1.0\columnwidth}
	\tiny
	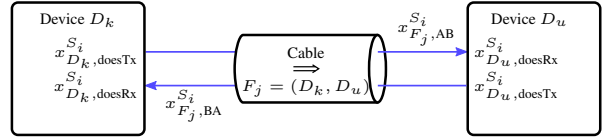
	\caption{Relation of the signal path encoding variables of a signal $S_i$ with the elements of a network}.
	\label{fig:routing:coupling}
\end{figure}
If a device $D_k$ emits a signal $S_i$, then a cable must transmit it away from the device. This is expressed with eq. \ref{eq:routing:txCable}. A cable $F_j = \left(D_k, \ast \right)$ that starts at an emitting device $D_k$ carries the signal in the direction $A \rightarrow B$. A cable $F_j = \left(\ast, D_k \right)$ carries signals to device $D_k$ (direction $B \rightarrow A$). This relationship of the variables to the individual elements is outlined by fig. \ref{fig:routing:coupling}.

A signal $S_i$ is always routed over a cable towards the receiving device $D_k$. The directional relationships for receiving is revers compared to sending. Eq. \ref{eq:routing:rxCable} sums up the number of incoming path segments of the signal $S_i$ for the device $D_k$.
\begin{align}
\var{S_i}{D_k , \text{doesTx}} =&%
\smashoperator{\sum_{ F_j \in \left\{ \set F \mid F_j = \left(D_k, \ast \right) \right\}}} \var{S_i}{F_j , \text{AB}} \qquad + \qquad
\smashoperator{\sum_{ F_j \in \left\{ \set F \mid F_j = \left(\ast, D_k \right) \right\}}} \var{S_i}{F_j , \text{BA}}
\label{eq:routing:txCable}\\
\nonumber\\
\var{S_i}{D_k , \text{doesRx}} =&%
\smashoperator{\sum_{ F_j \in \left\{ \set F \mid F_j = \left(D_k, \ast \right) \right\}}} \var{S_i}{F_j , \text{BA}} \qquad + \qquad
\smashoperator{\sum_{ F_j \in \left\{ \set F \mid F_j = \left(\ast, D_k \right) \right\}}} \var{S_i}{F_j , \text{AB}}
\label{eq:routing:rxCable}
\end{align}

Both equations restrict, in addition, that at most one outgoing and one incoming transmission of the signal exist. Multicasting is prohibited. The equations also prevent a signal from passing the same device twice.

\subsubsection{Signal-core-coupling}
\label{sec:algo:routing:resourceCable}
The type-dependent core-count limits the routing.
The two count variables $\var*{F_j}{useAB}$ and $\var*{F_j}{useBA}$ for each cable $F_j \in \set F$ count the number of transmitted signals in each direction.
Signal routing is directly coupled with the cable selection and indirectly with the device selection.
\begin{align}
\var*{F_j}{useAB} =& \sum_{ S_i \in \set S} \var{S_i}{F_j , \text{AB}}\\
\nonumber\\
\var*{F_j}{useBA} =& \sum_{ S_i \in \set S} \var{S_i}{F_j , \text{BA}}
\end{align}

The total number of transmitted signals for each cable $F_j \in \set F$ is calculated by adding up the individual transmission path variables $\var{S_i}{F_j , \text{AB}}$ and $\var{S_i}{F_j , \text{BA}}$ of each signal $S_i$ while taking into account the transmission direction.

\subsection{Attenuation}
\label{sec:algo:attenuation}

This work puts a special focus on including signal attenuation in the topology optimization problem.
The additional variables listed in table \ref{tbl:attenuation:vars} are added for attenuation computation.
These represent the power levels in all devices and cables for each signal $S_i \in \set S$. An exception is the variable $\var{S_i}{D_k , \text{opaqueRx}}$. 
\begin{table}[htbp]
	
	\centering
	\scriptsize
	\begin{tabular}{@{}llll@{}}
		\toprule
		\textbf{Designation} &
		\textbf{Symbol} &
		\textbf{Type} &
		\textbf{Exists for}
		\\	
		\midrule
		\multicolumn{4}{c}{\textbf{device-related variables}}\\
		
		Received input power & $\var{S_i}{D_k , \text{Rx}}$ & \reals & $\forall S_i \in \set S$, $\forall D_k \in \set D$ \\
		Opaque output power & $\var{S_i}{D_k , \text{transmit}}$ & \reals & $\forall S_i \in \set S$, $\forall D_k \in \set D$ \\
		Translucent output power & $\var{S_i}{D_k , \text{TxAvail}}$ & \reals & $\forall S_i \in \set S$, $\forall D_k \in \set D$ \\
		Used output power & $\var{S_i}{D_k , \text{Tx}}$ & \reals & $\forall S_i \in \set S$, $\forall D_k \in \set D$ \\
		Is opaque signal receiver & $\var{S_i}{D_k , \text{opaqueRx}}$ & \binary & $\forall S_i \in \set S$, $\forall D_k \in \set D$ \\
		
		\midrule
		\multicolumn{4}{c}{\textbf{cable-related variables}}\\
		
		Power at cable end $A B$ & $\var{S_i}{F_j , \text{powerAB}}$ & \reals & $\forall S_i \in \set S$, $\forall F_j \in \set F$ \\
		Power at cable end $B A$ & $\var{S_i}{F_j , \text{powerBA}}$ & \reals & $\forall S_i \in \set S$, $\forall F_j \in \set F$ \\
		
		\bottomrule
	\end{tabular}
	\caption{Additional variables required for attenuation}
	\label{tbl:attenuation:vars}
\end{table}

\subsubsection{Transmit Power Limits}
\label{sec:algo:attenuation:powerlimit}

A fully connected network with $\abs{\set D}$ devices is assumed. The longest possible signal path in it passes all $\abs{\set D}$ devices. Only translucent intermediate devices are of interest for the power considerations. Opaque devices can repower the signal.
Between each pair of devices, the signal is transmitted via a cable. In total, this results in $\abs{\set D} - 1$ cables in the signal path assuming that each device is not passed more than once.
The signal power at the receiving device is calculated by adding the transmit power to all attenuations occurring along the signal path. Conversely, the transmit power can be calculated starting from the receive power. Inserting maximum and minimum attenuation values yields the extreme conditions. 
A positive attenuation value is signal amplification. Maximum attenuation correlates with the smallest attenuation value. The minimum attenuation or gain correlates with the largest attenuation value.

With eq. \ref{eq:algo:attenuation:maxPower} a general limit is derived. All power values occurring in the network have to be within in the interval $\left[-P_{lim}, P_{lim} \right]$.
\begin{align}
&P_{lim} =
	P_{lim,Tx} +
	\left( \abs{\set D} - \num{2} \right) \cdot P_{lim,\Delta_{\set D}} \nonumber \\ &+
	\left( \abs{\set D} - \num{1} \right) \cdot P_{lim,\Delta_{\set F}} +
	P_{lim,Rx} \geq 0
\label{eq:algo:attenuation:maxPower}
\end{align}


The four terms necessary for attenuation are calculated from the maximum and minimum values of the cable and device types.

First, $P_{lim,Tx}$ denotes the largest absolute transmit power. For the optimization only the largest maximum as well as the smallest minimum transmit power $\vec{\hat{\tau}}{\set D}{\overlineText{Tx}}$ and $\vec{\check{\tau}}{\set D}{\underlineText{Tx}}$ are required. Since 
$\vec{\check{\tau}}{\set D}{\underlineText{Tx}} \leq
\vec{\check{\tau}}{\set D}{\overlineText{Tx}} \leq
\vec{\hat{\tau}}{\set D}{\overlineText{Tx}} $
and
$\vec{\check{\tau}}{\set D}{\underlineText{Tx}} \leq
\vec{\hat{\tau}}{\set D}{\underlineText{Tx}} \leq
\vec{\hat{\tau}}{\set D}{\overlineText{Tx}} $
, $\vec{\hat{\tau}}{\set D}{\underlineText{Tx}}$ and $\vec{\check{\tau}}{\set D}{\overlineText{Tx}} $ are eliminated, i.e. the largest minimum and smallest maximum transmit power are removed:
\begin{align}
	P_{lim,Tx} :=& \max \vec*{
			\abs{ \vec{\check{\tau}}{\set D}{\underlineText{Tx}} },\ 
			\abs{ \vec{\hat{\tau}}{\set D}{\overlineText{Tx}} } 
		}
	\label{sec:algo:attenuation:maxPower:Tx}
\end{align}

In the same way, $P_{lim,Rx}$ represents the largest received power. Again, the calculation requires only the largest maximum and smallest minimum power $\vec{\hat{\tau}}{\set D}{\overlineText{Rx}}$ and $\vec{\check{\tau}}{\set D}{\underlineText{Rx}}$. Similar to the receive power 
$\vec{\check{\tau}}{\set D}{\underlineText{Rx}} \leq
\vec{\check{\tau}}{\set D}{\overlineText{Rx}} \leq
\vec{\hat{\tau}}{\set D}{\overlineText{Rx}} $
and
$\vec{\check{\tau}}{\set D}{\underlineText{Rx}} \leq
\vec{\hat{\tau}}{\set D}{\underlineText{Rx}} \leq
\vec{\hat{\tau}}{\set D}{\overlineText{Rx}} $,
the smallest maximum as well as largest minimum received power $\vec{\check{\tau}}{\set D}{\overlineText{Rx}} $ and $\vec{\hat{\tau}}{\set D}{\underlineText{Rx}}$ do not need to be considered.%
\begin{align}
P_{lim,Rx} :=& \max \vec*{
	\abs{ \vec{\check{\tau}}{\set D}{\underlineText{Rx}} },\ 
	\abs{ \vec{\hat{\tau}}{\set D}{\overlineText{Rx}} } 
}
\label{sec:algo:attenuation:maxPower:Rx}
\end{align}

The remaining two terms are determined similarly. $P_{lim,\Delta_{\set D}}$ is determined by the largest absolute value of the internal device attenuation. The maximum absolute value of the cable-side transmission loss is used for the $P_{lim,\Delta_{\set F}}$.%
\begin{align}
P_{lim,\Delta_{\set D}} :=& \max \vec*{
	\abs{ \vec{\check{\tau}}{\set D}{\Delta} },\ 
	\abs{ \vec{\hat{\tau}}{\set D}{\Delta} } 
}
\label{sec:algo:attenuation:maxPower:DampingIntern}
\\
P_{lim,\Delta_{\set F}} :=& \max \vec*{
	\abs{ \vec{\check{\tau}}{\set F}{\Delta} },\ 
	\abs{ \vec{\hat{\tau}}{\set F}{\Delta} } 
}
\label{sec:algo:attenuation:maxPower:DampingFiber}
\end{align}

\subsubsection{Cable attenuation}

The power calculation is carried out according to an adapted calculation rule derived from \cite{FoaOpticalFiber} and \cite{FoaLoss}.
Along the signal path, starting with the transmitted power, all attenuation influences are accumulated. The largest attenuations are scattering and absorption effects in cables and devices as well as mechanical inaccuracies in the alignment (s. sec. \ref{sec:attenuation}). The latter is referred to as insertion loss at connectors.
The insertion loss is not assigned to the devices, but is considered as a part of the cable design.

All attenuations along a cable $F_j \in \set F$ are assumed to be summable to a single attenuation value $\property{F_j}{\Delta}$. This is possible since the length of each cable is known from the installation space dimensions. Furthermore, it is assumed that the plug-in or splice connections are included in this attenuation value.

\begin{figure}[ht]
	\centering
	\tiny
	\def\svgwidth{1.0\columnwidth}
	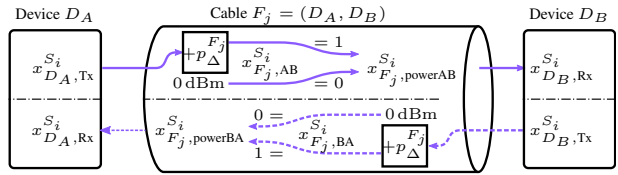
	\caption[Power flow along a cable $F_j$ for a signal $S_i$.]{Power flow along a cable $F_j = \vec*{D_A, D_B}$ for a signal $S_i$. The upper half of the figure shows how the power information propagates in the cable direction. The power flow against the direction is shown in the lower half of the figure.}
	\label{fig:attenuation:cableIntern}
\end{figure}

The basic idea of the power calculation is to add up the segment’s individual attenuations. 
However, since no signal path information is available when setting up the MILP constraints, all possible paths have to be considered. This leads to the model shown in fig. \ref{fig:attenuation:cableIntern}.

For each signal $S_i \in \set S$ and each cable $F_j \in \set F$ a set of constraints is defined, which calculates the power change along cable $F_j$ depending on the signal path. The path as well as the powers depend on the direction.

If the signal $S_i$ is transmitted as depicted in fig. \ref{fig:attenuation:cableIntern} from device $D_A$ to device $D_B$, then 
the signal power at the cable’s end, $\var{S_i}{F_j , \text{powerAB}}$, is obtained by adding the cable attenuation $\property{F_j}{\Delta}$ to the transmit power $\var{S_i}{D_A, \text{Tx}}$ of $D_A$ to be determined. The attenuation value is negative if signal attenuation occurs. Therefore, the attenuation has to be added. 

If the signal is not transmitted over that link, then $\var{S_i}{F_j , \text{powerAB}} = \SI{0}{\dBm}$ is applied to the power at the cable’s end. This choice is reasonable for the attenuation consideration in the devices. For the opposite direction $B \rightarrow A$ the conditions apply analogously. 
Formally, these conditions can be described mathematically as follows. $\var{S_i}{F_j , \text{AB}}$ is the transmission state.

\begin{align}
\var{S_i}{F_j , \text{powerAB}} =& \begin{cases}
\var{S_i}{D_A, \text{Tx}} + \property{F_j}{\Delta} & \text{if}\ \var{S_i}{F_j , \text{AB}} = \num{1}\\
\num{0} & \text{if}\ \var{S_i}{F_j , \text{AB}} = \num{0}
\end{cases}
\label{eq:algo:attenuation:cable:impl1}
\\%
\var{S_i}{F_j , \text{powerBA}} =& \begin{cases}
\var{S_i}{D_B, \text{Tx}} + \property{F_j}{\Delta} & \text{if}\ \var{S_i}{F_j , \text{BA}} = \num{1}\\
\num{0} & \text{if}\ \var{S_i}{F_j , \text{BA}} = \num{0}
\end{cases}
\label{eq:algo:attenuation:cable:impl2}
\end{align}

These distinctions are incompatible with the MILP constraints.
Eight inequalities are needed for a MILP compatible notation using the Big-M method \cite[p.~2]{indicatorConstraints}. In each case, the transmission state of the respective direction $\var{S_i}{F_j , \text{AB}}$ or $\var{S_i}{F_j , \text{BA}}$ serves as the decision variable. Since power is accounted, the power limit $P_{lim}$ serves as the Big-M constant.
\begin{align}
-\bigMexpr{\var{S_i}{F_j , \text{AB}}} & \leq \var{S_i}{F_j , \text{powerAB}} - \var{S_i}{D_A, \text{Tx}} + \property{F_j}{\Delta} \nonumber\\
                                       & \leq \bigMexpr{\var{S_i}{F_j , \text{AB}}}\\
-\bigMexpr*{\var{S_i}{F_j , \text{AB}}} & \leq \var{S_i}{F_j , \text{powerAB}} \leq \bigMexpr*{\var{S_i}{F_j , \text{AB}}}\\
\nonumber\\%
-\bigMexpr{\var{S_i}{F_j , \text{BA}}} & \leq \var{S_i}{F_j , \text{powerBA}} - \var{S_i}{D_B, \text{Tx}} + \property{F_j}{\Delta} \nonumber\\
                                       & \leq \bigMexpr{\var{S_i}{F_j , \text{BA}}}\\
-\bigMexpr*{\var{S_i}{F_j , \text{BA}}} & \leq \var{S_i}{F_j , \text{powerBA}} \leq \bigMexpr*{\var{S_i}{F_j , \text{BA}}}
\end{align}

\subsection{Attenuation Consideration in Devices}

Attenuation behaves different for devices. Power calculations there are more complex as for cables. The challenge is to model opaque and translucent devices by a common set of constraints. Separation is not possible due to the common type model.

\begin{figure}[htbp]
	\centering
	\tiny
	\def\svgwidth{1.0\columnwidth}
	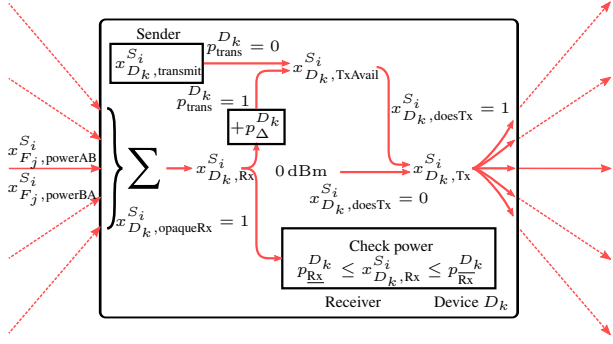
	\caption{Power flow within a device $D_k$ for a signal $S_i$. The mixture of opaque and translucent devices creates complexity within the attenuation consideration.}
	\label{fig:attenuation:deviceInternal}
\end{figure}

\subsubsection{Received Power}
Fig. \ref{fig:attenuation:deviceInternal} shows how power flows through a device $D_k \in \set D$. Assuming that at most one cable $F_j$ carries the signal $S_i \in \set S$ to the device, the incoming signal strength is calculated by summation. For this purpose the power at the cable’s end is set to \SI{0}{\dBm} when no signal is transmitted. This incoming signal power $\var{S_i}{D_k , \text{Rx}}$ corresponds to the power at the end of the cable transmitting to the device:
\begin{equation}
\var{S_i}{D_k , \text{Rx}} = %
\smashoperator{\sum_{F_j \in \left\{ \set F \mid F_j = \left(\ast, D_k \right) \right\}}} \var{S_i}{F_j , \text{powerAB}} \qquad + \qquad
\smashoperator{\sum_{F_j \in \left\{ \set F \mid F_j = \left(D_k, \ast \right) \right\}}} \var{S_i}{F_j , \text{powerBA}}
\end{equation}

The auxiliary variable $\var{S_i}{D_k , \text{opaqueRx}}$ encodes whether an opaque type is assigned to the associated device $D_k$ and the signal $S_i$ is received by the device. It is determined from combination of the receive state $\var{S_i}{D_k , \text{doesRx}}$ and the complementary value of the device property $\property*{D_k}{trans}$.

According to the general encoding of logical AND conditions \cite[4~ff]{FormulateILP} this results in MILP :

\begin{align}
\var{S_i}{D_k , \text{opaqueRx}} &\leq \mathbin{\phantom{(}} \var{S_i}{D_k , \text{doesRx}}\\ %
\var{S_i}{D_k , \text{opaqueRx}} &\leq \binaryNot{\property*{D_k}{trans}}\\ %
\var{S_i}{D_k , \text{opaqueRx}} &\geq \binaryNot{\property*{D_k}{trans}} + \var{S_i}{D_k , \text{doesRx}} - \num{1} %
\end{align}

If the device $D_k$ is opaque, then the received power should neither overpower the sensor ($\property{D_k}{\overlineText{Rx}}$) nor be below its sensitivity $\property{D_k}{\underlineText{Rx}}$. Formally, however, it is considered whether the signal $S_i$ is received at all. Indeed, if there is no reception, then in the received power $\var{S_i}{D_k , \text{Rx}} = 0$ is meaningless and the sensitivity is not considered. All incoming signal powers are zero, thus their sum is also zero. A check is performed if the device $D_k$ is opaque and the signal $S_i$ is received. This condition is just represented by the auxiliary variable $\var{S_i}{D_k , \text{opaqueRx}}$. Formally, the implication below is obtained with the decision variable $\var{S_i}{D_k , \text{opaqueRx}}$:%
\begin{equation}
\var{S_i}{D_k , \text{opaqueRx}} = \num{1} \quad\implies\quad%
\property{D_k}{\underlineText{Rx}} \leq \var{S_i}{D_k , \text{Rx}} \leq \property{D_k}{\overlineText{Rx}}
\end{equation}

As before, this is represented with the Big-M notation \cite[p.~2]{indicatorConstraints}. Again, the power limit $P_{lim}$ is used as the Big-M constant.
\begin{align}
\var{S_i}{D_k , \text{Rx}} - \property{D_k}{\overlineText{Rx}} \quad&\leq\quad \bigMexpr{\var{S_i}{D_k , \text{opaqueRx}}}\\
-\var{S_i}{D_k , \text{Rx}} + \property{D_k}{\underlineText{Rx}} \quad&\leq\quad \bigMexpr{\var{S_i}{D_k , \text{opaqueRx}}}
\end{align}

It should be noted that attenuations occurring during reception in the device are not modeled. If an optical path to the sensor inside the opaque device degrades the signal, this attenuation is added to the sensitivity range in advance during the type declaration. Further possibly desired margins are to be added in advance to the declaration.

\subsubsection{Signal Dispatch and Internal Attenuation}

The power calculation of the signal dispatch requires several steps. First, the theoretically available transmit power $\var{S_i}{D_k , \text{TxAvail}}$ is determined.
In the case of opaque devices, this corresponds directly to the actual transmitter power $\var{S_i}{D_k , \text{transmit}}$. For translucent devices, the available transmit power is calculated from the received power $\var{S_i}{D_k , \text{Rx}}$ and the internal attenuation $\property{D_k}{\Delta}$ of the device:
\begin{equation}
\var{S_i}{D_k , \text{TxAvail}} = \begin{cases}
	\var{S_i}{D_k , \text{Rx}} + \property{D_k}{\Delta} & \text{if}\ \property*{D_k}{trans} = \num{1}\\
	\var{S_i}{D_k , \text{transmit}} & \text{if}\ \property*{D_k}{trans} = \num{0}
\end{cases}
\label{eq:algo:attenuation:txAvailCase}
\end{equation}

The decision variable $\property*{D_k}{trans} = \num{1}$ encodes whether an opaque or translucent device is present. At this point, the device attenuation of translucent devices is included. For opaque devices there is no consideration of attenuation values, these (if present) are accounted for with the transmitter power as part of the device type declaration.
The inclusion of the case distinction (\ref{eq:algo:attenuation:txAvailCase}) is based again on Big-M:
\begin{align}	
&-\bigMexpr{\property*{D_k}{trans}} \leq%
\var{S_i}{D_k , \text{TxAvail}} - \var{S_i}{D_k , \text{Rx}} - \property{D_k}{\Delta} \nonumber \\
& \leq \bigMexpr{\property*{D_k}{trans}}
\\%
&-\bigMexpr*{\property*{D_k}{trans}} \leq%
\var{S_i}{D_k , \text{TxAvail}} - \var{S_i}{D_k , \text{transmit}} %
\leq \bigMexpr*{\property*{D_k}{trans}}
\end{align}

The transmitter power $\var{S_i}{D_k , \text{transmit}}$ is freely selectable within a power range $\left[ \property{D_k}{\underlineText{Tx}},\ \property{D_k}{\overlineText{Tx}} \right]$. Consequently, transmitters with adjustable power can be considered in the optimization. A distinction between opaque and translucent devices is not made. For the latter, the range is set to $\left\{\num{0}\right\}$.
Moreover, an additional constraint is required:
\begin{equation}
\property{D_k}{\underlineText{Tx}} \leq	\var{S_i}{D_k , \text{transmit}} \leq \property{D_k}{\overlineText{Tx}}
\end{equation}

The available transmit power $\var{S_i}{D_k , \text{TxAvail}}$ is calculated independently of the fact whether the device $D_k$ transmits the signal $S_i$ at all. The latter is stored in the status variable $\var{S_i}{D_k , \text{doesTx}}$. If $S_i$ is not transmitted, a transmit power of \SI{0}{\dBm} is desired. A case differentiation is introduced for calculating the actual transmitted signal power $\var{S_i}{D_k , \text{Tx}}$:
\begin{equation}
\var{S_i}{D_k , \text{Tx}} = \begin{cases}
	\var{S_i}{D_k , \text{TxAvail}} & \text{if}\ \var{S_i}{D_k , \text{doesTx}} = \num{1}\\
	\num{0} & \text{if}\ \var{S_i}{D_k , \text{doesTx}} = \num{0}
\end{cases}
\label{eq:algo:attenuation:txCase}
\end{equation}

When the signal $S_i$ is transmitted, the actual transmit power $\var{S_i}{D_k , \text{Tx}}$ is equal to the available power $\var{S_i}{D_k , \text{TxAvail}}$. If no transmission takes place, the transmit power is zero. For an overview of what actual transmit powers $\var{S_i}{D_k , \text{Tx}}$ are assumed for the different cases see table \ref{tbl:attenuation:deviceInternal}.
\begin{table}[htbp]
	\centering

	\begin{tabular}{@{}lccc@{}}
		\toprule
		\textbf{State} &
		$\property*{D_k}{trans}$ &
		$\var{S_i}{D_k , \text{doesTx}}$ &
		$\var{S_i}{D_k , \text{Tx}}$
		\\	
		\midrule
		Opaque, not transmitting & 0 & 0 & \SI{0}{\dBm} \\
		Translucent, no signal & 1 & 0 & \SI{0}{\dBm} \\
		\midrule
		Opaque, transmitting & 0 & 1 & $\var{S_i}{D_k , \text{transmit}}$ \\
		Translucent, with signal & 1 & 1 & $\var{S_i}{D_k , \text{Rx}} + \property{D_k}{\Delta}$ \\
		
		\bottomrule
	\end{tabular}
	
	\caption{The actual transmit power considered for the possible states of the translucent property and transmit status of a signal $S_i$ in device $D_k$}
	\label{tbl:attenuation:deviceInternal}
\end{table}
The case discrimination (\ref{eq:algo:attenuation:txCase}) is again represented by four constraints using the Big-M method. The Big-M constant must not be exceeded or undercut by the absolute power bound $P_{lim}$ used as a limit.
\begin{align}	
	&-\bigMexpr{\var{S_i}{D_k , \text{doesTx}}} \leq%
		\var{S_i}{D_k , \text{Tx}} - \var{S_i}{D_k , \text{TxAvail}} \nonumber \\
	&\leq \bigMexpr{\var{S_i}{D_k , \text{doesTx}}}
	\\%
	&-\bigMexpr*{\var{S_i}{D_k , \text{doesTx}}} \leq%
		\var{S_i}{D_k , \text{Tx}} \nonumber \\
	&\leq \bigMexpr*{\var{S_i}{D_k , \text{doesTx}}}
\end{align}

%% file: Figures/3-5-3_kopplung.pdf_tex
\begingroup%
  \makeatletter%
  \providecommand\color[2][]{%
    \errmessage{(Inkscape) Color is used for the text in Inkscape, but the package 'color.sty' is not loaded}%
    \renewcommand\color[2][]{}%
  }%
  \providecommand\transparent[1]{%
    \errmessage{(Inkscape) Transparency is used (non-zero) for the text in Inkscape, but the package 'transparent.sty' is not loaded}%
    \renewcommand\transparent[1]{}%
  }%
  \providecommand\rotatebox[2]{#2}%
  \newcommand*\fsize{\dimexpr\f@size pt\relax}%
  \newcommand*\lineheight[1]{\fontsize{\fsize}{#1\fsize}\selectfont}%
  \ifx\svgwidth\undefined%
    \setlength{\unitlength}{360bp}%
    \ifx\svgscale\undefined%
      \relax%
    \else%
      \setlength{\unitlength}{\unitlength * \real{\svgscale}}%
    \fi%
  \else%
    \setlength{\unitlength}{\svgwidth}%
  \fi%
  \global\let\svgwidth\undefined%
  \global\let\svgscale\undefined%
  \makeatother%
  \begin{picture}(1,0.27777778)%
    \lineheight{1}%
    \setlength\tabcolsep{0pt}%
    \put(0,0){\includegraphics[width=\unitlength,page=1]{3-5-3_kopplung.pdf}}%
    \put(0.22366937,0.15793861){\color[rgb]{0,0,0}\makebox(0,0)[rt]{\lineheight{1.25}\smash{\begin{tabular}[t]{r}$\var{S_i}{D_k , \text{doesTx}}$\end{tabular}}}}%
    \put(0.12491044,0.21331516){\color[rgb]{0,0,0}\makebox(0,0)[t]{\lineheight{1.25}\smash{\begin{tabular}[t]{c}Device $D_k$\\\end{tabular}}}}%
    \put(0.86796594,0.21331516){\color[rgb]{0,0,0}\makebox(0,0)[t]{\lineheight{1.25}\smash{\begin{tabular}[t]{c}Device $D_u$\\\end{tabular}}}}%
    \put(0.77668859,0.15793861){\color[rgb]{0,0,0}\makebox(0,0)[lt]{\lineheight{1.25}\smash{\begin{tabular}[t]{l}$\var{S_i}{D_u , \text{doesRx}}$\end{tabular}}}}%
    \put(0.65046519,0.19973207){\color[rgb]{0,0,0}\makebox(0,0)[lt]{\lineheight{1.25}\smash{\begin{tabular}[t]{l}$\var{S_i}{F_j , \text{AB}}$\end{tabular}}}}%
    \put(0.36325947,0.07370878){\color[rgb]{0,0,0}\makebox(0,0)[rt]{\lineheight{1.25}\smash{\begin{tabular}[t]{r}$\var{S_i}{F_j , \text{BA}}$\end{tabular}}}}%
    \put(0.50017904,0.10238316){\color[rgb]{0,0,0}\makebox(0,0)[t]{\lineheight{1.25}\smash{\begin{tabular}[t]{c}$F_j = \left(D_k, D_u\right)$\\\end{tabular}}}}%
    \put(0.49996265,0.15590955){\color[rgb]{0,0,0}\makebox(0,0)[t]{\lineheight{1.25}\smash{\begin{tabular}[t]{c}Cable\\\end{tabular}}}}%
    \put(0.22366937,0.10238304){\color[rgb]{0,0,0}\makebox(0,0)[rt]{\lineheight{1.25}\smash{\begin{tabular}[t]{r}$\var{S_i}{D_k , \text{doesRx}}$\end{tabular}}}}%
    \put(0.77668859,0.10238304){\color[rgb]{0,0,0}\makebox(0,0)[lt]{\lineheight{1.25}\smash{\begin{tabular}[t]{l}$\var{S_i}{D_u , \text{doesTx}}$\end{tabular}}}}%
    \put(0,0){\includegraphics[width=\unitlength,page=2]{3-5-3_kopplung.pdf}}%
  \end{picture}%
\endgroup%

%% file: Figures/3-5-4_attenuation_cable.pdf_tex
\begingroup%
  \makeatletter%
  \providecommand\color[2][]{%
    \errmessage{(Inkscape) Color is used for the text in Inkscape, but the package 'color.sty' is not loaded}%
    \renewcommand\color[2][]{}%
  }%
  \providecommand\transparent[1]{%
    \errmessage{(Inkscape) Transparency is used (non-zero) for the text in Inkscape, but the package 'transparent.sty' is not loaded}%
    \renewcommand\transparent[1]{}%
  }%
  \providecommand\rotatebox[2]{#2}%
  \newcommand*\fsize{\dimexpr\f@size pt\relax}%
  \newcommand*\lineheight[1]{\fontsize{\fsize}{#1\fsize}\selectfont}%
  \ifx\svgwidth\undefined%
    \setlength{\unitlength}{390bp}%
    \ifx\svgscale\undefined%
      \relax%
    \else%
      \setlength{\unitlength}{\unitlength * \real{\svgscale}}%
    \fi%
  \else%
    \setlength{\unitlength}{\svgwidth}%
  \fi%
  \global\let\svgwidth\undefined%
  \global\let\svgscale\undefined%
  \makeatother%
  \begin{picture}(1,0.30769231)%
    \lineheight{1}%
    \setlength\tabcolsep{0pt}%
    \put(0,0){\includegraphics[width=\unitlength,page=1]{3-5-4_attenuation_cable.pdf}}%
    \put(0.14316282,0.17143062){\color[rgb]{0,0,0}\makebox(0,0)[rt]{\lineheight{1.25}\smash{\begin{tabular}[t]{r}$\var{S_i}{D_A , \text{Tx}}$\end{tabular}}}}%
    \put(0.07764173,0.26100889){\color[rgb]{0,0,0}\makebox(0,0)[t]{\lineheight{1.25}\smash{\begin{tabular}[t]{c}Device $D_A$\\\end{tabular}}}}%
    \put(0.92379564,0.26100889){\color[rgb]{0,0,0}\makebox(0,0)[t]{\lineheight{1.25}\smash{\begin{tabular}[t]{c}Device $D_B$\\\end{tabular}}}}%
    \put(0.59030842,0.17154762){\color[rgb]{0,0,0}\makebox(0,0)[lt]{\lineheight{1.25}\smash{\begin{tabular}[t]{l}$\var{S_i}{F_j , \text{powerAB}}$\end{tabular}}}}%
    \put(0.47790024,0.26117414){\color[rgb]{0,0,0}\makebox(0,0)[t]{\lineheight{1.25}\smash{\begin{tabular}[t]{c}Cable $F_j = \left(D_A, D_B\right)$\\\end{tabular}}}}%
    \put(0.5029009,0.14578953){\color[rgb]{0,0,0}\makebox(0,0)[lt]{\lineheight{1.25}\smash{\begin{tabular}[t]{l}$= \num{0}$\end{tabular}}}}%
    \put(0,0){\includegraphics[width=\unitlength,page=2]{3-5-4_attenuation_cable.pdf}}%
    \put(0.5029009,0.21353822){\color[rgb]{0,0,0}\makebox(0,0)[lt]{\lineheight{1.25}\smash{\begin{tabular}[t]{l}$= \num{1}$\end{tabular}}}}%
    \put(0.32458436,0.19589976){\color[rgb]{0,0,0}\makebox(0,0)[t]{\lineheight{1.25}\smash{\begin{tabular}[t]{c}$+\property{F_j}{\Delta}$\end{tabular}}}}%
    \put(0,0){\includegraphics[width=\unitlength,page=3]{3-5-4_attenuation_cable.pdf}}%
    \put(0.35727009,0.14654702){\color[rgb]{0,0,0}\makebox(0,0)[rt]{\lineheight{1.25}\smash{\begin{tabular}[t]{r}\SI{0}{\dBm}\\\end{tabular}}}}%
    \put(0.47960109,0.18113643){\color[rgb]{0,0,0}\makebox(0,0)[rt]{\lineheight{1.25}\smash{\begin{tabular}[t]{r}$\var{S_i}{F_j , \text{AB}}$\end{tabular}}}}%
    \put(0,0){\includegraphics[width=\unitlength,page=4]{3-5-4_attenuation_cable.pdf}}%
    \put(0.62077612,0.09526496){\color[rgb]{0,0,0}\makebox(0,0)[lt]{\lineheight{1.25}\smash{\begin{tabular}[t]{l}\SI{0}{\dBm}\\\end{tabular}}}}%
    \put(0.47725987,0.06265899){\color[rgb]{0,0,0}\makebox(0,0)[lt]{\lineheight{1.25}\smash{\begin{tabular}[t]{l}$\var{S_i}{F_j , \text{BA}}$\end{tabular}}}}%
    \put(0.45396006,0.09450747){\color[rgb]{0,0,0}\makebox(0,0)[rt]{\lineheight{1.25}\smash{\begin{tabular}[t]{r}$\num{0} =$\end{tabular}}}}%
    \put(0.45396006,0.03100825){\color[rgb]{0,0,0}\makebox(0,0)[rt]{\lineheight{1.25}\smash{\begin{tabular}[t]{r}$\num{1} =$\end{tabular}}}}%
    \put(0.6529927,0.04205358){\color[rgb]{0,0,0}\makebox(0,0)[t]{\lineheight{1.25}\smash{\begin{tabular}[t]{c}$+\property{F_j}{\Delta}$\end{tabular}}}}%
    \put(0.85877035,0.0688665){\color[rgb]{0,0,0}\makebox(0,0)[lt]{\lineheight{1.25}\smash{\begin{tabular}[t]{l}$\var{S_i}{D_B, \text{Tx}}$\end{tabular}}}}%
    \put(0.85877035,0.17143062){\color[rgb]{0,0,0}\makebox(0,0)[lt]{\lineheight{1.25}\smash{\begin{tabular}[t]{l}$\var{S_i}{D_B , \text{Rx}}$\end{tabular}}}}%
    \put(0.14316279,0.0688665){\color[rgb]{0,0,0}\makebox(0,0)[rt]{\lineheight{1.25}\smash{\begin{tabular}[t]{r}$\var{S_i}{D_A , \text{Rx}}$\end{tabular}}}}%
    \put(0.39050482,0.0689835){\color[rgb]{0,0,0}\makebox(0,0)[rt]{\lineheight{1.25}\smash{\begin{tabular}[t]{r}$\var{S_i}{F_j , \text{powerBA}}$\end{tabular}}}}%
  \end{picture}%
\endgroup%

%% file: Figures/3-5-4_attenuation.pdf_tex
\begingroup%
  \makeatletter%
  \providecommand\color[2][]{%
    \errmessage{(Inkscape) Color is used for the text in Inkscape, but the package 'color.sty' is not loaded}%
    \renewcommand\color[2][]{}%
  }%
  \providecommand\transparent[1]{%
    \errmessage{(Inkscape) Transparency is used (non-zero) for the text in Inkscape, but the package 'transparent.sty' is not loaded}%
    \renewcommand\transparent[1]{}%
  }%
  \providecommand\rotatebox[2]{#2}%
  \newcommand*\fsize{\dimexpr\f@size pt\relax}%
  \newcommand*\lineheight[1]{\fontsize{\fsize}{#1\fsize}\selectfont}%
  \ifx\svgwidth\undefined%
    \setlength{\unitlength}{405bp}%
    \ifx\svgscale\undefined%
      \relax%
    \else%
      \setlength{\unitlength}{\unitlength * \real{\svgscale}}%
    \fi%
  \else%
    \setlength{\unitlength}{\svgwidth}%
  \fi%
  \global\let\svgwidth\undefined%
  \global\let\svgscale\undefined%
  \makeatother%
  \begin{picture}(1,0.56790123)%
    \lineheight{1}%
    \setlength\tabcolsep{0pt}%
    \put(0,0){\includegraphics[width=\unitlength,page=1]{3-5-4_attenuation.pdf}}%
    \put(0.57282832,0.05633328){\color[rgb]{0,0,0}\makebox(0,0)[t]{\lineheight{1.25}\smash{\begin{tabular}[t]{c}Receiver\\\end{tabular}}}}%
    \put(0.67078371,0.27609956){\color[rgb]{0,0,0}\makebox(0,0)[lt]{\lineheight{1.25}\smash{\begin{tabular}[t]{l}$\var{S_i}{D_k , \text{Tx}}$\end{tabular}}}}%
    \put(0.25230371,0.4963192){\color[rgb]{0,0,0}\makebox(0,0)[t]{\lineheight{1.25}\smash{\begin{tabular}[t]{c}Sender\end{tabular}}}}%
    \put(0.53077237,0.27114797){\color[rgb]{0,0,0}\makebox(0,0)[rt]{\lineheight{1.25}\smash{\begin{tabular}[t]{r}\SI{0}{\dBm}\end{tabular}}}}%
    \put(0.4732528,0.44276654){\color[rgb]{0,0,0}\makebox(0,0)[lt]{\lineheight{1.25}\smash{\begin{tabular}[t]{l}$\var{S_i}{D_k , \text{TxAvail}}$\end{tabular}}}}%
    \put(0.25301322,0.44893904){\color[rgb]{0,0,0}\makebox(0,0)[t]{\lineheight{1.25}\smash{\begin{tabular}[t]{c}$\var{S_i}{D_k , \text{transmit}}$\end{tabular}}}}%
    \put(0.00411699,0.30560971){\color[rgb]{0,0,0}\makebox(0,0)[lt]{\lineheight{1.25}\smash{\begin{tabular}[t]{l}$\var{S_i}{F_j , \text{powerAB}}$\end{tabular}}}}%
    \put(0.63574921,0.14729114){\color[rgb]{0,0,0}\makebox(0,0)[t]{\lineheight{1.25}\smash{\begin{tabular}[t]{c}Check power\\\end{tabular}}}}%
    \put(0.6357295,0.10956511){\color[rgb]{0,0,0}\makebox(0,0)[t]{\lineheight{1.25}\smash{\begin{tabular}[t]{c}$\property{D_k}{\underlineText{Rx}} \leq \var{S_i}{D_k , \text{Rx}} \leq \property{D_k}{\overlineText{Rx}}$\end{tabular}}}}%
    \put(0,0){\includegraphics[width=\unitlength,page=2]{3-5-4_attenuation.pdf}}%
    \put(0.31384644,0.27619247){\color[rgb]{0,0,0}\makebox(0,0)[lt]{\lineheight{1.25}\smash{\begin{tabular}[t]{l}$\var{S_i}{D_k , \text{Rx}}$\\\end{tabular}}}}%
    \put(0,0){\includegraphics[width=\unitlength,page=3]{3-5-4_attenuation.pdf}}%
    \put(0.39182665,0.47670347){\color[rgb]{0,0,0}\makebox(0,0)[t]{\lineheight{1.25}\smash{\begin{tabular}[t]{c}$\property*{D_k}{trans} = \num{0}$\end{tabular}}}}%
    \put(0.40314402,0.38734331){\color[rgb]{0,0,0}\makebox(0,0)[rt]{\lineheight{1.25}\smash{\begin{tabular}[t]{r}$\property*{D_k}{trans} = \num{1}$\end{tabular}}}}%
    \put(0.40252072,0.18977267){\color[rgb]{0,0,0}\makebox(0,0)[rt]{\lineheight{1.25}\smash{\begin{tabular}[t]{r}$\var{S_i}{D_k , \text{opaqueRx}} = \num{1}$\end{tabular}}}}%
    \put(0,0){\includegraphics[width=\unitlength,page=4]{3-5-4_attenuation.pdf}}%
    \put(0.63483411,0.37495782){\color[rgb]{0,0,0}\makebox(0,0)[lt]{\lineheight{1.25}\smash{\begin{tabular}[t]{l}$\var{S_i}{D_k , \text{doesTx}} = \num{1}$\end{tabular}}}}%
    \put(0.59869227,0.22768602){\color[rgb]{0,0,0}\makebox(0,0)[t]{\lineheight{1.25}\smash{\begin{tabular}[t]{c}$\var{S_i}{D_k , \text{doesTx}} = \num{0}$\end{tabular}}}}%
    \put(0,0){\includegraphics[width=\unitlength,page=5]{3-5-4_attenuation.pdf}}%
    \put(0.41373915,0.34446801){\color[rgb]{0,0,0}\makebox(0,0)[t]{\lineheight{1.25}\smash{\begin{tabular}[t]{c}$+\property{D_k}{\Delta}$\end{tabular}}}}%
    \put(0.83100018,0.0560148){\color[rgb]{0,0,0}\makebox(0,0)[rt]{\lineheight{1.25}\smash{\begin{tabular}[t]{r}Device $D_k$\\\end{tabular}}}}%
    \put(0.00411699,0.24514838){\color[rgb]{0,0,0}\makebox(0,0)[lt]{\lineheight{1.25}\smash{\begin{tabular}[t]{l}$\var{S_i}{F_j , \text{powerBA}}$\end{tabular}}}}%
    \put(0,0){\includegraphics[width=\unitlength,page=6]{3-5-4_attenuation.pdf}}%
  \end{picture}%
\endgroup%

%% file: validation.tex
\section{Validation}
\label{sec:validation}

Validation is performed with small scenarios each dedicated to a certain aspect of the optimization approach. 

The optimization is implemented in Python. The input and output data are modeled with the Open Avionics Architecture Model\footnote{\url{http://www.oaam.de}} (OAAM) \cite{annighoefer2019c}. OAAM is a domain-specific model developed to represent the logical and physical architecture of cyber-physical aircraft systems and the underlying avionics architecture. It includes elements to model the possible installation space, signal needs, and the component types with their properties used in the optimization. The required information is read from OAAM with Python and transformed to the matrix representation of the MILP. The result of the optimization is converted back into model elements and stored in OAAM. The MILP problem is solved using Gurobi 8 on an Intel Core i3-2100 with 8 GB RAM. 

\begin{figure}[htb] 
	\centering
		\def\svgwidth{\linewidth}
		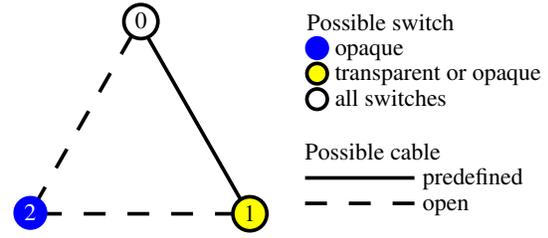
		\caption{Validation model A}
		\label{fig:valid:preA}

\end{figure}

Two basic models are used for validation experiments:

Model A as depicted in fig. \ref{fig:valid:preA} defines three switches as well as a fixed cable and type between devices 0 and 1. All other cables are subject to the optimization.
For the optimization, opaque and translucent switches (s. tab. \ref{tbl:valid:deviceTypesStd}) are available. Switch 2 is fixed as an opaque device.
Two cable types (s. tab. \ref{tbl:valid:cableTypesStd}) are defined: The unidirectional two-core cable type as well as the bidirectional three-core cable type. The predefined cable instance between switch~0 and switch~1 is unidirectional.

\begin{figure}[htb] 
	\centering
	
		\def\svgwidth{\linewidth}
		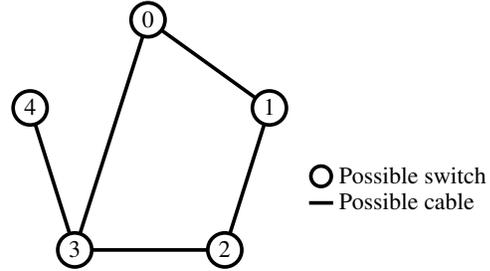
		\caption{Validation model B}
		\label{fig:valid:preB}

\end{figure}

Model B is a network consisting of five switches \numrange{0}{4}. Five possible cables are declared as depicted in fig. \ref{fig:valid:preB}.
The switch types are again opaque and translucent. Three bidirectional cable types are defined: A single-core cable type with high attenuation as well as two- and three-core bidirectional types (s. tab. \ref{tbl:valid:cableTypesStd}). There is no restriction on the type assignment for switches and cables.

All scenarios use cost minimization as an objective function, i.e.

\begin{align}
	\text{minimize} \quad \sum_{D_k \in \set D} \property*{D_k}{cost} + \sum_{F_j \in \set F} \property*{F_j}{cost}. \label{eq:cost:cost}
\end{align}

\begin{table}[htbp]
	\centering
	\tiny
	\setlength{\tabcolsep}{3pt}
	\begin{tabular}{@{}Xlccccccccl@{}}
		\toprule
		& \multicolumn{9}{c}{\textbf{Properties}} \\
		\textbf{Types}
		& $\vec{\tau}{\set D}{\text{ports}_t}$
		& $\vec{\tau}{\set D}{\Delta_t}$
		& $\vec{\tau}{\set D}{\underlineText{Rx}_t}$
		& $\vec{\tau}{\set D}{\overlineText{Rx}_t}$
		& $\vec{\tau}{\set D}{\underlineText{Tx}_t}$
		& $\vec{\tau}{\set D}{\overlineText{Tx}_t}$
		& $\vec{\tau}{\set D}{\text{trans}_t}$
		& $\vec{\tau}{\set D}{\text{cost}_t}$
		& Model
		\\
		\midrule
		
		opaque
		& \num{4}
		& ---
		& \SI{-14}{}
		& \SI{0.5}{}
		& \SI{-5}{}
		& \SI{0}{}
		& \num{0}
		& \num{300}
		& A,B
		\\

		translucent 
		& \num{2}
		& \SI{-0.5}{\decibel}
		& ---
		& ---
		& ---
		& ---
		& \num{1}
		& \num{100}
		& A,B
		\\
		
		\bottomrule
	\end{tabular}
	
	\caption{Properties of the switch types for validation. Entries with --- are modeled with the value zero. RX and TX values are in \dBm.}
	\label{tbl:valid:deviceTypesStd}
\end{table}

\begin{table}[htbp]
	\centering
	\scriptsize
	\setlength{\tabcolsep}{3pt}
	\begin{tabular}{@{}Xlcccll@{}}
		\toprule
		& \multicolumn{4}{c}{\textbf{Properties}} \\
		\textbf{Types}
		& $\vec{\tau}{\set F}{\text{cores}_t}$
		& $\vec{\tau}{\set F}{\Delta_t}$
		& $\vec{\tau}{\set F}{\text{cost}_t}$
		& Direction
		& Model
		\\
		\midrule

		Bidirectional, 1 core
		& \num{1}
		& \SI{-15.0}{\decibel}
		& \num{1}
		& \num{0} (Bidirectional)
		& B
		\\
		
		Bidirectional, 2 cores
		& \num{2}
		& \phantom{1}\SI{-2.0}{\decibel}
		& \num{30}
		& \num{0} (Bidirectional)
		& B
		\\
		
		Unidirectional, 2 cores
		& \num{2}
		& \phantom{1}\SI{-2.0}{\decibel}
		& \num{30}
		& \num{3} (Unidirectional)
		& A
		\\
		
		Bidirectional, 3 cores
		& \num{3}
		& \phantom{1}\SI{-2.0}{\decibel}
		& \num{50}
		& \num{0} (Bidirectional)		
		& A,B
		\\
		
		\bottomrule
	\end{tabular}
	
	\caption{Properties of cable types for validation. Attenuation includes connectors and internal cable losses}
	\label{tbl:valid:cableTypesStd}
\end{table}

\subsection{Scenario 1: Unidirectional Cables}

\begin{figure}[htbp] 
	\centering
	
	\fontsize{8.5pt}{8.5pt}
	\def\svgwidth{0.9\linewidth}
	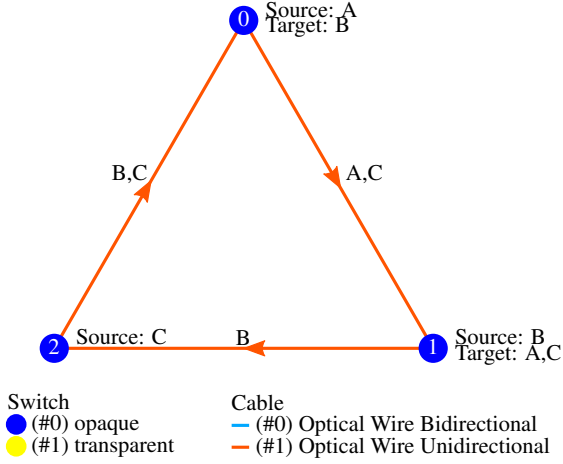

	\caption{Optimized topology of scenario 1}
	\label{fig:valid:resA}
\end{figure}

The first scenario is based on model~A with three defined signals identified by letters A, B and C. Fig. \ref{fig:valid:resA} depicts optimized network topology. The colors of the devices and cables indicate the associated types, while the letters at the connection illustrate the respective signal path.

All three switches are assigned to the opaque switch type. Each transmits and receives at least one signal. All connections are optimized to use unidirectional cables. Their defined transmission directions are indicated by arrows.
The cable connection between devices 0 and 1 is predefined in the model, but its direction was undefined.
A cycle formed by the cables can be observed. This corresponds to the signals to be transmitted. Only one of the two signals A and B can be transmitted via the cable between devices 0 and 1 due to directionality. The signal that is not transmitted via this cable, in this case B, has to take a detour via switch 2. Signal C fits within the second available core of the unidirectional cables in each case.
The two generated cables (0--2 and 1--2) must exist because of the required paths for signal B and C, but the type is not specified. Since there is no need to use the more expensive bidirectional cable (cost: \num{50}), the less expensive unidirectional cable (cost: \num{30}) is chosen. Therefore, this topology has the lowest possible cost under the given constraints. 

\subsubsection{Scenario 2: Attenuation}

\begin{figure}[htbp] 
	\centering
	
	\fontsize{8.5pt}{8.5pt}
	\def\svgwidth{0.9\linewidth}
	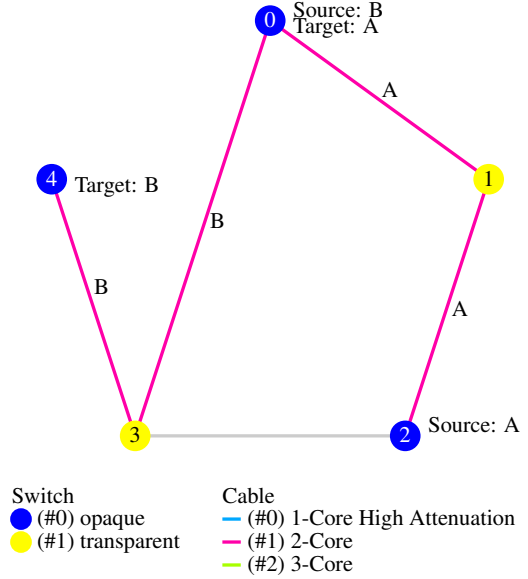
	\caption{Optimized topology of scenario 2}
	\label{fig:valid:resB:2}
\end{figure}

As a second scenario, the topology of model B with two signals A and B is optimized. The network cost serves as the objective function.
Two opaque and translucent switch types per position are possible. The translucent switch can connect at most two cables in the validation.
The three bidirectional cable types are selectable. The single-core cable type has a high attenuation.
Sources and destinations of signals A and B can be taken from the optimized topology in fig. \ref{fig:valid:resB:2}.

As expected, all source and destination switches are opaque switches in the optimized topology. 
Instead of the (intentionally) inexpensive single-core cable type the significantly costlier dual-core cable type is used, even though it is not needed due to the number of signals. The reason is the attenuation over the signal paths. This is verified using signal A as an example. This results in the following calculation for the cable type used:

\begin{align*}
	\property{D_2}{\overlineText{Tx}} + \property{\left( D_1, D_2 \right)}{\Delta} + \property{D_1}{\Delta} + \property{\left( D_0, D_1 \right)}{\Delta} &= \\
	 \phantom{-}\SI{0}{\dBm} - \SI{2}{\decibel} - \SI{0.5}{\decibel} - \SI{2}{\decibel} &= \\
\SI{-4.5}{\dBm}\\
	\property{D_2}{\underlineText{Tx}} + \property{\left( D_1, D_2 \right)}{\Delta} + \property{D_1}{\Delta} + \property{\left( D_0, D_1 \right)}{\Delta} & = \\
	\SI{-5}{\dBm} - \SI{2}{\decibel} - \SI{0.5}{\decibel} - \SI{2}{\decibel} & = \\
\SI{-9.5}{\dBm}
\end{align*}

These values are within the permissible range of \SIrange{-14}{0.5}{\dBm} of the opaque receiver. If only one cable is exchanged for the one with poorer attenuation, the receiver cannot receive the signal:

\begin{align*}
\property{D_2}{\overlineText{Tx}} + \property{\left( D_1, D_2 \right)}{\Delta} + \property{D_1}{\Delta} + \property{\left( D_0, D_1 \right)}{\Delta} & = \\
\phantom{-}\SI{0}{\dBm} - \SI{15}{\decibel} - \SI{0.5}{\decibel} - \SI{2}{\decibel} & = \\
\SI{-17.5}{\dBm}\\
\property{D_2}{\underlineText{Tx}} + \property{\left( D_1, D_2 \right)}{\Delta} + \property{D_1}{\Delta} + \property{\left( D_0, D_1 \right)}{\Delta} & = \\
\SI{-5}{\dBm} - \SI{15}{\decibel} - \SI{0.5}{\decibel} - \SI{2}{\decibel} & = \\
\SI{-22.5}{\dBm}
\end{align*}

Consequently, the cable type with poor attenuation is not possible. In addition, a common signal routing via switch 3 is not possible, because the translucent switch 3 allows at most two connections. It would have to be replaced by an opaque switch for a more compact architecture, but this comes with overall higher cost. Instead of the two translucent switches and one cable (cost:  \num{230}), one opaque switch would be needed (cost: \num{300}). The result is, thus, minimized in cost and all signal power levels are legal.

\subsubsection{Scenario 3: Core Count}

\begin{figure}[htbp] 
	\centering
	
	\fontsize{8.5pt}{8.5pt}
	\def\svgwidth{0.9\linewidth}
	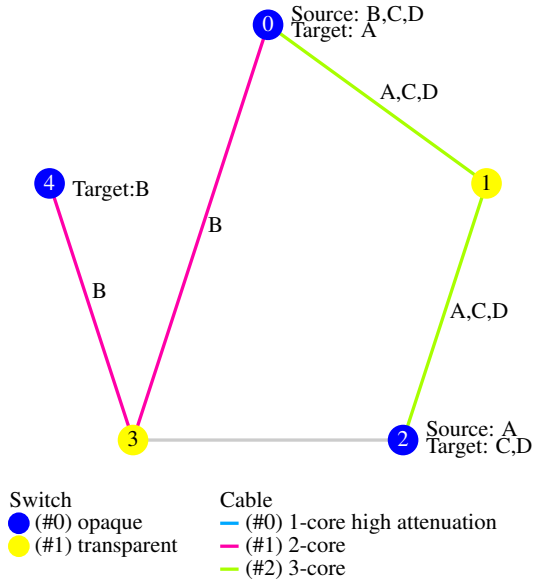
	\caption{Optimized topology of scenario 3. Compared to scenario 2, two cables have a higher core count.}
	\label{fig:valid:resB:4}
\end{figure}

Scenario 3 extends scenario 2. In addition, two more signals C and D are defined. Both start at switch 0 and have switch 2 as their destination. This requires two more cores.
The result of the optimization is shown in fig. \ref{fig:valid:resB:4}. It can be seen that the path of signal B is unchanged. Also, the switch and cable types along its path are the same.
However, a change of the cable types for the other signal paths is necessary. Between switches 0, 1, and 2, the three signals A, C and D are transmitted. This makes the third cable type with three cores mandatory.

\subsubsection{Scenario 4: Core count vs. Switch Ports}
\begin{figure}[htbp] 
	\centering
	
	\fontsize{8.5pt}{8.5pt}
	\def\svgwidth{0.9\linewidth}
	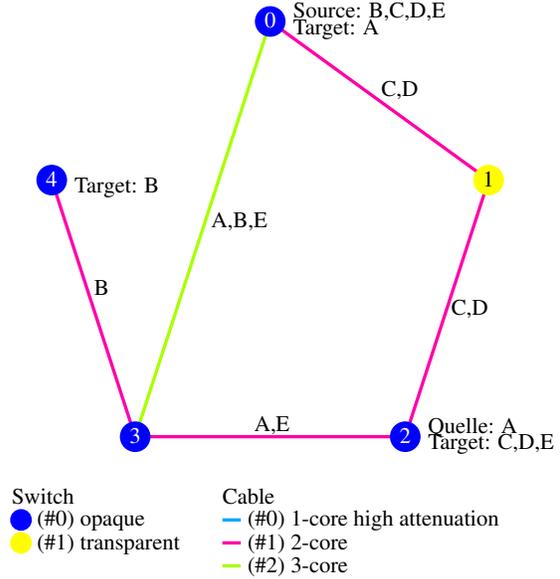
	\caption{Optimized topology of scenario 4 adding a fifth signal.}
	\label{fig:valid:resB:5}
\end{figure}

Scenario 4 extends scenario 3 with the additional signal E. E has the same target and destination as the signals C and D. This time, there is a significant change in the topology, which can be seen in fig. \ref{fig:valid:resB:5}.
The four signals A, C, D, and E require at least two connections to switch 2. The available cable types are limited to a maximum of three cores.
Compared to scenario 3, one additional core from switch 0 to 2 is necessary. This results in the change of switch 3 that is mandatory. Instead of the previous translucent switch type, the opaque type is assigned, because of the required connection capacities. The other switch types remain optimal according to the requirements. Considering the cables, the two path options (0-3-2 and 0-1-2) have both two segments. Instead of introducing one upgraded and one additional 3-core cable to the solution of scenario 2 (cost +70), only one is necessary (cost -20). Therefore, the result is the cheapest possible topology.

\subsubsection{Scenario 5: Free interconnection}

As a last validation scenario, a further variation of the third scenario is optimized. The network is now automatically completed by the algorithm, i.e. no cable routes are predefined. All cables can be created.

\begin{figure}[htbp] 
	\centering
	
	\fontsize{8.5pt}{8.5pt}
	\def\svgwidth{0.9\linewidth}
	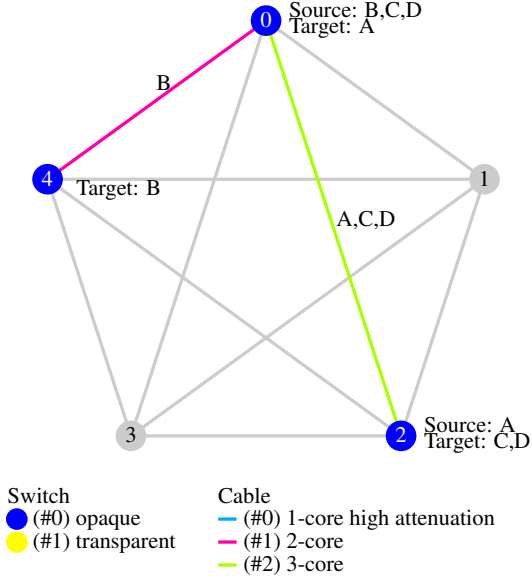
	\caption{Optimized topology of scenario 5}
	\label{fig:valid:resB:4F}
\end{figure}

The resulting topology shown in fig. \ref{fig:valid:resB:4F} corresponds to the expectation. A smaller topology is not achievable with the given specifications, because the start and target switch are fixed by the signals. The result corresponds to the minimum required number of cables with the most reasonable cable types in terms of cost and attenuation.

%% file: Figures/features/preOptimierung.pdf_tex
\begingroup%
  \makeatletter%
  \providecommand\color[2][]{%
    \errmessage{(Inkscape) Color is used for the text in Inkscape, but the package 'color.sty' is not loaded}%
    \renewcommand\color[2][]{}%
  }%
  \providecommand\transparent[1]{%
    \errmessage{(Inkscape) Transparency is used (non-zero) for the text in Inkscape, but the package 'transparent.sty' is not loaded}%
    \renewcommand\transparent[1]{}%
  }%
  \providecommand\rotatebox[2]{#2}%
  \newcommand*\fsize{\dimexpr\f@size pt\relax}%
  \newcommand*\lineheight[1]{\fontsize{\fsize}{#1\fsize}\selectfont}%
  \ifx\svgwidth\undefined%
    \setlength{\unitlength}{375bp}%
    \ifx\svgscale\undefined%
      \relax%
    \else%
      \setlength{\unitlength}{\unitlength * \real{\svgscale}}%
    \fi%
  \else%
    \setlength{\unitlength}{\svgwidth}%
  \fi%
  \global\let\svgwidth\undefined%
  \global\let\svgscale\undefined%
  \makeatother%
  \begin{picture}(1,0.4)%
    \lineheight{1}%
    \setlength\tabcolsep{0pt}%
    \put(0,0){\includegraphics[width=\unitlength,page=1]{preOptimierung.pdf}}%
    \put(0.22987153,0.34581232){\color[rgb]{0,0,0}\makebox(0,0)[t]{\lineheight{1.25}\smash{\begin{tabular}[t]{c}0\end{tabular}}}}%
    \put(0,0){\includegraphics[width=\unitlength,page=2]{preOptimierung.pdf}}%
    \put(0.40913883,0.02933334){\color[rgb]{0.04705882,0.04705882,0.04705882}\makebox(0,0)[t]{\lineheight{1.25}\smash{\begin{tabular}[t]{c}1\end{tabular}}}}%
    \put(0,0){\includegraphics[width=\unitlength,page=3]{preOptimierung.pdf}}%
    \put(0.048,0.03082314){\color[rgb]{1,1,1}\makebox(0,0)[t]{\lineheight{1.25}\smash{\begin{tabular}[t]{c}2\end{tabular}}}}%
    \put(0.50300051,0.34320033){\color[rgb]{0,0,0}\makebox(0,0)[lt]{\lineheight{1.25}\smash{\begin{tabular}[t]{l}Possible switch\end{tabular}}}}%
    \put(0,0){\includegraphics[width=\unitlength,page=4]{preOptimierung.pdf}}%
    \put(0.54920051,0.30260033){\color[rgb]{0,0,0}\makebox(0,0)[lt]{\lineheight{1.25}\smash{\begin{tabular}[t]{l}opaque\end{tabular}}}}%
    \put(0,0){\includegraphics[width=\unitlength,page=5]{preOptimierung.pdf}}%
    \put(0.54920051,0.26060033){\color[rgb]{0,0,0}\makebox(0,0)[lt]{\lineheight{1.25}\smash{\begin{tabular}[t]{l}transparent or opaque\end{tabular}}}}%
    \put(0,0){\includegraphics[width=\unitlength,page=6]{preOptimierung.pdf}}%
    \put(0.54920047,0.21859938){\color[rgb]{0,0,0}\makebox(0,0)[lt]{\lineheight{1.25}\smash{\begin{tabular}[t]{l}all switches\end{tabular}}}}%
    \put(0.5,0.12953984){\color[rgb]{0,0,0}\makebox(0,0)[lt]{\lineheight{1.25}\smash{\begin{tabular}[t]{l}Possible cable\end{tabular}}}}%
    \put(0,0){\includegraphics[width=\unitlength,page=7]{preOptimierung.pdf}}%
    \put(0.69349496,0.08893984){\color[rgb]{0,0,0}\makebox(0,0)[lt]{\lineheight{1.25}\smash{\begin{tabular}[t]{l}predefined\end{tabular}}}}%
    \put(0,0){\includegraphics[width=\unitlength,page=8]{preOptimierung.pdf}}%
    \put(0.69349496,0.04693984){\color[rgb]{0,0,0}\makebox(0,0)[lt]{\lineheight{1.25}\smash{\begin{tabular}[t]{l}open\end{tabular}}}}%
  \end{picture}%
\endgroup%

%% file: Figures/validation/preOptimierung.pdf_tex
\begingroup%
  \makeatletter%
  \providecommand\color[2][]{%
    \errmessage{(Inkscape) Color is used for the text in Inkscape, but the package 'color.sty' is not loaded}%
    \renewcommand\color[2][]{}%
  }%
  \providecommand\transparent[1]{%
    \errmessage{(Inkscape) Transparency is used (non-zero) for the text in Inkscape, but the package 'transparent.sty' is not loaded}%
    \renewcommand\transparent[1]{}%
  }%
  \providecommand\rotatebox[2]{#2}%
  \newcommand*\fsize{\dimexpr\f@size pt\relax}%
  \newcommand*\lineheight[1]{\fontsize{\fsize}{#1\fsize}\selectfont}%
  \ifx\svgwidth\undefined%
    \setlength{\unitlength}{375bp}%
    \ifx\svgscale\undefined%
      \relax%
    \else%
      \setlength{\unitlength}{\unitlength * \real{\svgscale}}%
    \fi%
  \else%
    \setlength{\unitlength}{\svgwidth}%
  \fi%
  \global\let\svgwidth\undefined%
  \global\let\svgscale\undefined%
  \makeatother%
  \begin{picture}(1,0.5)%
    \lineheight{1}%
    \setlength\tabcolsep{0pt}%
    \put(0,0){\includegraphics[width=\unitlength,page=1]{preOptimierung.pdf}}%
    \put(0.23501553,0.42781661){\color[rgb]{0,0,0}\makebox(0,0)[t]{\lineheight{1.25}\smash{\begin{tabular}[t]{c}0\end{tabular}}}}%
    \put(0,0){\includegraphics[width=\unitlength,page=2]{preOptimierung.pdf}}%
    \put(0.43473741,0.28271017){\color[rgb]{0,0,0}\makebox(0,0)[t]{\lineheight{1.25}\smash{\begin{tabular}[t]{c}1\end{tabular}}}}%
    \put(0,0){\includegraphics[width=\unitlength,page=3]{preOptimierung.pdf}}%
    \put(0.3613036,0.04885009){\color[rgb]{0,0,0}\makebox(0,0)[t]{\lineheight{1.25}\smash{\begin{tabular}[t]{c}2\end{tabular}}}}%
    \put(0,0){\includegraphics[width=\unitlength,page=4]{preOptimierung.pdf}}%
    \put(0.11443381,0.04885009){\color[rgb]{0,0,0}\makebox(0,0)[t]{\lineheight{1.25}\smash{\begin{tabular}[t]{c}3\end{tabular}}}}%
    \put(0,0){\includegraphics[width=\unitlength,page=5]{preOptimierung.pdf}}%
    \put(0.041,0.28271017){\color[rgb]{0,0,0}\makebox(0,0)[t]{\lineheight{1.25}\smash{\begin{tabular}[t]{c}4\end{tabular}}}}%
    \put(0,0){\includegraphics[width=\unitlength,page=6]{preOptimierung.pdf}}%
    \put(0.5492,0.17060045){\color[rgb]{0,0,0}\makebox(0,0)[lt]{\lineheight{1.25}\smash{\begin{tabular}[t]{l}Possible switch\end{tabular}}}}%
    \put(0,0){\includegraphics[width=\unitlength,page=7]{preOptimierung.pdf}}%
    \put(0.54920084,0.12675937){\color[rgb]{0,0,0}\makebox(0,0)[lt]{\lineheight{1.25}\smash{\begin{tabular}[t]{l}Possible cable\end{tabular}}}}%
  \end{picture}%
\endgroup%

%% file: Figures/optimiert.pdf_tex
\begingroup%
  \makeatletter%
  \providecommand\color[2][]{%
    \errmessage{(Inkscape) Color is used for the text in Inkscape, but the package 'color.sty' is not loaded}%
    \renewcommand\color[2][]{}%
  }%
  \providecommand\transparent[1]{%
    \errmessage{(Inkscape) Transparency is used (non-zero) for the text in Inkscape, but the package 'transparent.sty' is not loaded}%
    \renewcommand\transparent[1]{}%
  }%
  \providecommand\rotatebox[2]{#2}%
  \newcommand*\fsize{\dimexpr\f@size pt\relax}%
  \newcommand*\lineheight[1]{\fontsize{\fsize}{#1\fsize}\selectfont}%
  \ifx\svgwidth\undefined%
    \setlength{\unitlength}{393.75bp}%
    \ifx\svgscale\undefined%
      \relax%
    \else%
      \setlength{\unitlength}{\unitlength * \real{\svgscale}}%
    \fi%
  \else%
    \setlength{\unitlength}{\svgwidth}%
  \fi%
  \global\let\svgwidth\undefined%
  \global\let\svgscale\undefined%
  \makeatother%
  \begin{picture}(1,0.85714286)%
    \lineheight{1}%
    \setlength\tabcolsep{0pt}%
    \put(0,0){\includegraphics[width=\unitlength,page=1]{optimiert.pdf}}%
    \put(0.62810879,0.50591925){\color[rgb]{0,0,0}\makebox(0,0)[lt]{\lineheight{1.25}\smash{\begin{tabular}[t]{l}A,C\end{tabular}}}}%
    \put(0,0){\includegraphics[width=\unitlength,page=2]{optimiert.pdf}}%
    \put(0.26502773,0.50591952){\color[rgb]{0,0,0}\makebox(0,0)[rt]{\lineheight{1.25}\smash{\begin{tabular}[t]{r}B,C\end{tabular}}}}%
    \put(0,0){\includegraphics[width=\unitlength,page=3]{optimiert.pdf}}%
    \put(0.4405068,0.20484){\color[rgb]{0,0,0}\makebox(0,0)[t]{\lineheight{1.25}\smash{\begin{tabular}[t]{c}B\end{tabular}}}}%
    \put(0,0){\includegraphics[width=\unitlength,page=4]{optimiert.pdf}}%
    \put(0.44115242,0.78698346){\color[rgb]{1,1,1}\makebox(0,0)[t]{\lineheight{1.25}\smash{\begin{tabular}[t]{c}0\end{tabular}}}}%
    \put(0.48115236,0.77624273){\color[rgb]{0,0,0}\makebox(0,0)[lt]{\lineheight{1.25}\smash{\begin{tabular}[t]{l}Target: B\end{tabular}}}}%
    \put(0.48115236,0.80513158){\color[rgb]{0,0,0}\makebox(0,0)[lt]{\lineheight{1.25}\smash{\begin{tabular}[t]{l}Source: A\end{tabular}}}}%
    \put(0,0){\includegraphics[width=\unitlength,page=5]{optimiert.pdf}}%
    \put(0.78756208,0.18698432){\color[rgb]{1,1,1}\makebox(0,0)[t]{\lineheight{1.25}\smash{\begin{tabular}[t]{c}1\end{tabular}}}}%
    \put(0.82756204,0.17624358){\color[rgb]{0,0,0}\makebox(0,0)[lt]{\lineheight{1.25}\smash{\begin{tabular}[t]{l}Target: A,C\end{tabular}}}}%
    \put(0.82756204,0.20513242){\color[rgb]{0,0,0}\makebox(0,0)[lt]{\lineheight{1.25}\smash{\begin{tabular}[t]{l}Source: B\end{tabular}}}}%
    \put(0,0){\includegraphics[width=\unitlength,page=6]{optimiert.pdf}}%
    \put(0.09474276,0.18698432){\color[rgb]{1,1,1}\makebox(0,0)[t]{\lineheight{1.25}\smash{\begin{tabular}[t]{c}2\end{tabular}}}}%
    \put(0.13474271,0.20513242){\color[rgb]{0,0,0}\makebox(0,0)[lt]{\lineheight{1.25}\smash{\begin{tabular}[t]{l}Source: C\end{tabular}}}}%
    \put(0.00913713,0.08685804){\color[rgb]{0,0,0}\makebox(0,0)[lt]{\lineheight{1.25}\smash{\begin{tabular}[t]{l}Switch\end{tabular}}}}%
    \put(0,0){\includegraphics[width=\unitlength,page=7]{optimiert.pdf}}%
    \put(0.05313706,0.04819142){\color[rgb]{0,0,0}\makebox(0,0)[lt]{\lineheight{1.25}\smash{\begin{tabular}[t]{l}(\#0) opaque\end{tabular}}}}%
    \put(0,0){\includegraphics[width=\unitlength,page=8]{optimiert.pdf}}%
    \put(0.05313706,0.00819148){\color[rgb]{0,0,0}\makebox(0,0)[lt]{\lineheight{1.25}\smash{\begin{tabular}[t]{l}(\#1) transparent\end{tabular}}}}%
    \put(0.41904702,0.08510506){\color[rgb]{0,0,0}\makebox(0,0)[lt]{\lineheight{1.25}\smash{\begin{tabular}[t]{l}Cable\end{tabular}}}}%
    \put(0,0){\includegraphics[width=\unitlength,page=9]{optimiert.pdf}}%
    \put(0.46304695,0.04643845){\color[rgb]{0,0,0}\makebox(0,0)[lt]{\lineheight{1.25}\smash{\begin{tabular}[t]{l}(\#0) Optical Wire Bidirectional\end{tabular}}}}%
    \put(0,0){\includegraphics[width=\unitlength,page=10]{optimiert.pdf}}%
    \put(0.46304695,0.00643851){\color[rgb]{0,0,0}\makebox(0,0)[lt]{\lineheight{1.25}\smash{\begin{tabular}[t]{l}(\#1) Optical Wire Unidirectional\end{tabular}}}}%
  \end{picture}%
\endgroup%

%% file: Figures/case2.pdf_tex
\begingroup%
  \makeatletter%
  \providecommand\color[2][]{%
    \errmessage{(Inkscape) Color is used for the text in Inkscape, but the package 'color.sty' is not loaded}%
    \renewcommand\color[2][]{}%
  }%
  \providecommand\transparent[1]{%
    \errmessage{(Inkscape) Transparency is used (non-zero) for the text in Inkscape, but the package 'transparent.sty' is not loaded}%
    \renewcommand\transparent[1]{}%
  }%
  \providecommand\rotatebox[2]{#2}%
  \newcommand*\fsize{\dimexpr\f@size pt\relax}%
  \newcommand*\lineheight[1]{\fontsize{\fsize}{#1\fsize}\selectfont}%
  \ifx\svgwidth\undefined%
    \setlength{\unitlength}{375bp}%
    \ifx\svgscale\undefined%
      \relax%
    \else%
      \setlength{\unitlength}{\unitlength * \real{\svgscale}}%
    \fi%
  \else%
    \setlength{\unitlength}{\svgwidth}%
  \fi%
  \global\let\svgwidth\undefined%
  \global\let\svgscale\undefined%
  \makeatother%
  \begin{picture}(1,1.1)%
    \lineheight{1}%
    \setlength\tabcolsep{0pt}%
    \put(0,0){\includegraphics[width=\unitlength,page=1]{case2.pdf}}%
    \put(0.69170378,0.87999999){\color[rgb]{0,0,0}\makebox(0,0)[lt]{\lineheight{1.25}\smash{\begin{tabular}[t]{l}A\end{tabular}}}}%
    \put(0,0){\includegraphics[width=\unitlength,page=2]{case2.pdf}}%
    \put(0.81970377,0.48){\color[rgb]{0,0,0}\makebox(0,0)[lt]{\lineheight{1.25}\smash{\begin{tabular}[t]{l}A\end{tabular}}}}%
    \put(0,0){\includegraphics[width=\unitlength,page=3]{case2.pdf}}%
    \put(0.16572134,0.52){\color[rgb]{0,0,0}\makebox(0,0)[lt]{\lineheight{1.25}\smash{\begin{tabular}[t]{l}B\end{tabular}}}}%
    \put(0,0){\includegraphics[width=\unitlength,page=4]{case2.pdf}}%
    \put(0.37772136,0.64){\color[rgb]{0,0,0}\makebox(0,0)[lt]{\lineheight{1.25}\smash{\begin{tabular}[t]{l}B\end{tabular}}}}%
    \put(0,0){\includegraphics[width=\unitlength,page=5]{case2.pdf}}%
    \put(0.48744375,1.00833333){\color[rgb]{1,1,1}\makebox(0,0)[t]{\lineheight{1.25}\smash{\begin{tabular}[t]{c}0\end{tabular}}}}%
    \put(0.52944376,0.99705554){\color[rgb]{0,0,0}\makebox(0,0)[lt]{\lineheight{1.25}\smash{\begin{tabular}[t]{l}Target: A\end{tabular}}}}%
    \put(0.52944376,1.02738889){\color[rgb]{0,0,0}\makebox(0,0)[lt]{\lineheight{1.25}\smash{\begin{tabular}[t]{l}Source: B\end{tabular}}}}%
    \put(0,0){\includegraphics[width=\unitlength,page=6]{case2.pdf}}%
    \put(0.88688752,0.71812048){\color[rgb]{0,0,0}\makebox(0,0)[t]{\lineheight{1.25}\smash{\begin{tabular}[t]{c}1\end{tabular}}}}%
    \put(0,0){\includegraphics[width=\unitlength,page=7]{case2.pdf}}%
    \put(0.73431355,0.2485462){\color[rgb]{1,1,1}\makebox(0,0)[t]{\lineheight{1.25}\smash{\begin{tabular}[t]{c}2\end{tabular}}}}%
    \put(0.77631355,0.26760176){\color[rgb]{0,0,0}\makebox(0,0)[lt]{\lineheight{1.25}\smash{\begin{tabular}[t]{l}Source: A\end{tabular}}}}%
    \put(0,0){\includegraphics[width=\unitlength,page=8]{case2.pdf}}%
    \put(0.24057397,0.2485462){\color[rgb]{0,0,0}\makebox(0,0)[t]{\lineheight{1.25}\smash{\begin{tabular}[t]{c}3\end{tabular}}}}%
    \put(0,0){\includegraphics[width=\unitlength,page=9]{case2.pdf}}%
    \put(0.08799998,0.71812048){\color[rgb]{1,1,1}\makebox(0,0)[t]{\lineheight{1.25}\smash{\begin{tabular}[t]{c}4\end{tabular}}}}%
    \put(0.13000003,0.70684269){\color[rgb]{0,0,0}\makebox(0,0)[lt]{\lineheight{1.25}\smash{\begin{tabular}[t]{l}Target: B\end{tabular}}}}%
    \put(0.015,0.137358){\color[rgb]{0,0,0}\makebox(0,0)[lt]{\lineheight{1.25}\smash{\begin{tabular}[t]{l}Switch\end{tabular}}}}%
    \put(0,0){\includegraphics[width=\unitlength,page=10]{case2.pdf}}%
    \put(0.0612,0.096758){\color[rgb]{0,0,0}\makebox(0,0)[lt]{\lineheight{1.25}\smash{\begin{tabular}[t]{l}(\#0) opaque\end{tabular}}}}%
    \put(0,0){\includegraphics[width=\unitlength,page=11]{case2.pdf}}%
    \put(0.0612,0.054758){\color[rgb]{0,0,0}\makebox(0,0)[lt]{\lineheight{1.25}\smash{\begin{tabular}[t]{l}(\#1) transparent\end{tabular}}}}%
    \put(0.40000001,0.13735938){\color[rgb]{0,0,0}\makebox(0,0)[lt]{\lineheight{1.25}\smash{\begin{tabular}[t]{l}Cable\end{tabular}}}}%
    \put(0,0){\includegraphics[width=\unitlength,page=12]{case2.pdf}}%
    \put(0.44620001,0.09675938){\color[rgb]{0,0,0}\makebox(0,0)[lt]{\lineheight{1.25}\smash{\begin{tabular}[t]{l}(\#0) 1-Core High Attenuation\end{tabular}}}}%
    \put(0,0){\includegraphics[width=\unitlength,page=13]{case2.pdf}}%
    \put(0.44620001,0.05475938){\color[rgb]{0,0,0}\makebox(0,0)[lt]{\lineheight{1.25}\smash{\begin{tabular}[t]{l}(\#1) 2-Core\end{tabular}}}}%
    \put(0,0){\includegraphics[width=\unitlength,page=14]{case2.pdf}}%
    \put(0.44620001,0.01275938){\color[rgb]{0,0,0}\makebox(0,0)[lt]{\lineheight{1.25}\smash{\begin{tabular}[t]{l}(\#2) 3-Core\end{tabular}}}}%
  \end{picture}%
\endgroup%

%% file: Figures/case4.pdf_tex
\begingroup%
  \makeatletter%
  \providecommand\color[2][]{%
    \errmessage{(Inkscape) Color is used for the text in Inkscape, but the package 'color.sty' is not loaded}%
    \renewcommand\color[2][]{}%
  }%
  \providecommand\transparent[1]{%
    \errmessage{(Inkscape) Transparency is used (non-zero) for the text in Inkscape, but the package 'transparent.sty' is not loaded}%
    \renewcommand\transparent[1]{}%
  }%
  \providecommand\rotatebox[2]{#2}%
  \newcommand*\fsize{\dimexpr\f@size pt\relax}%
  \newcommand*\lineheight[1]{\fontsize{\fsize}{#1\fsize}\selectfont}%
  \ifx\svgwidth\undefined%
    \setlength{\unitlength}{375bp}%
    \ifx\svgscale\undefined%
      \relax%
    \else%
      \setlength{\unitlength}{\unitlength * \real{\svgscale}}%
    \fi%
  \else%
    \setlength{\unitlength}{\svgwidth}%
  \fi%
  \global\let\svgwidth\undefined%
  \global\let\svgscale\undefined%
  \makeatother%
  \begin{picture}(1,1.1)%
    \lineheight{1}%
    \setlength\tabcolsep{0pt}%
    \put(0,0){\includegraphics[width=\unitlength,page=1]{case4.pdf}}%
    \put(0.69170378,0.88421552){\color[rgb]{0,0,0}\makebox(0,0)[lt]{\lineheight{1.25}\smash{\begin{tabular}[t]{l}A,C,D\end{tabular}}}}%
    \put(0,0){\includegraphics[width=\unitlength,page=2]{case4.pdf}}%
    \put(0.81970377,0.4842155){\color[rgb]{0,0,0}\makebox(0,0)[lt]{\lineheight{1.25}\smash{\begin{tabular}[t]{l}A,C,D\end{tabular}}}}%
    \put(0,0){\includegraphics[width=\unitlength,page=3]{case4.pdf}}%
    \put(0.16572134,0.52){\color[rgb]{0,0,0}\makebox(0,0)[lt]{\lineheight{1.25}\smash{\begin{tabular}[t]{l}B\end{tabular}}}}%
    \put(0,0){\includegraphics[width=\unitlength,page=4]{case4.pdf}}%
    \put(0.37772136,0.64){\color[rgb]{0,0,0}\makebox(0,0)[lt]{\lineheight{1.25}\smash{\begin{tabular}[t]{l}B\end{tabular}}}}%
    \put(0,0){\includegraphics[width=\unitlength,page=5]{case4.pdf}}%
    \put(0.48744375,1.00833333){\color[rgb]{1,1,1}\makebox(0,0)[t]{\lineheight{1.25}\smash{\begin{tabular}[t]{c}0\end{tabular}}}}%
    \put(0.52944376,0.99705554){\color[rgb]{0,0,0}\makebox(0,0)[lt]{\lineheight{1.25}\smash{\begin{tabular}[t]{l}Target: A\end{tabular}}}}%
    \put(0.52944376,1.02738889){\color[rgb]{0,0,0}\makebox(0,0)[lt]{\lineheight{1.25}\smash{\begin{tabular}[t]{l}Source: B,C,D\end{tabular}}}}%
    \put(0,0){\includegraphics[width=\unitlength,page=6]{case4.pdf}}%
    \put(0.88688752,0.71812048){\color[rgb]{0,0,0}\makebox(0,0)[t]{\lineheight{1.25}\smash{\begin{tabular}[t]{c}1\end{tabular}}}}%
    \put(0,0){\includegraphics[width=\unitlength,page=7]{case4.pdf}}%
    \put(0.73431355,0.2485462){\color[rgb]{1,1,1}\makebox(0,0)[t]{\lineheight{1.25}\smash{\begin{tabular}[t]{c}2\end{tabular}}}}%
    \put(0.77631355,0.23726842){\color[rgb]{0,0,0}\makebox(0,0)[lt]{\lineheight{1.25}\smash{\begin{tabular}[t]{l}Target: C,D\end{tabular}}}}%
    \put(0.77631355,0.26760176){\color[rgb]{0,0,0}\makebox(0,0)[lt]{\lineheight{1.25}\smash{\begin{tabular}[t]{l}Source: A\end{tabular}}}}%
    \put(0,0){\includegraphics[width=\unitlength,page=8]{case4.pdf}}%
    \put(0.24057397,0.2485462){\color[rgb]{0,0,0}\makebox(0,0)[t]{\lineheight{1.25}\smash{\begin{tabular}[t]{c}3\end{tabular}}}}%
    \put(0,0){\includegraphics[width=\unitlength,page=9]{case4.pdf}}%
    \put(0.08799998,0.71812048){\color[rgb]{1,1,1}\makebox(0,0)[t]{\lineheight{1.25}\smash{\begin{tabular}[t]{c}4\end{tabular}}}}%
    \put(0.13000003,0.70684269){\color[rgb]{0,0,0}\makebox(0,0)[lt]{\lineheight{1.25}\smash{\begin{tabular}[t]{l}Target:B\end{tabular}}}}%
    \put(0.015,0.137358){\color[rgb]{0,0,0}\makebox(0,0)[lt]{\lineheight{1.25}\smash{\begin{tabular}[t]{l}Switch\end{tabular}}}}%
    \put(0,0){\includegraphics[width=\unitlength,page=10]{case4.pdf}}%
    \put(0.0612,0.096758){\color[rgb]{0,0,0}\makebox(0,0)[lt]{\lineheight{1.25}\smash{\begin{tabular}[t]{l}(\#0) opaque\end{tabular}}}}%
    \put(0,0){\includegraphics[width=\unitlength,page=11]{case4.pdf}}%
    \put(0.0612,0.054758){\color[rgb]{0,0,0}\makebox(0,0)[lt]{\lineheight{1.25}\smash{\begin{tabular}[t]{l}(\#1) transparent\end{tabular}}}}%
    \put(0.40000002,0.13735938){\color[rgb]{0,0,0}\makebox(0,0)[lt]{\lineheight{1.25}\smash{\begin{tabular}[t]{l}Cable\end{tabular}}}}%
    \put(0,0){\includegraphics[width=\unitlength,page=12]{case4.pdf}}%
    \put(0.44620002,0.09675938){\color[rgb]{0,0,0}\makebox(0,0)[lt]{\lineheight{1.25}\smash{\begin{tabular}[t]{l}(\#0) 1-core high attenuation\end{tabular}}}}%
    \put(0,0){\includegraphics[width=\unitlength,page=13]{case4.pdf}}%
    \put(0.44620002,0.05475938){\color[rgb]{0,0,0}\makebox(0,0)[lt]{\lineheight{1.25}\smash{\begin{tabular}[t]{l}(\#1) 2-core\end{tabular}}}}%
    \put(0,0){\includegraphics[width=\unitlength,page=14]{case4.pdf}}%
    \put(0.44620002,0.01275938){\color[rgb]{0,0,0}\makebox(0,0)[lt]{\lineheight{1.25}\smash{\begin{tabular}[t]{l}(\#2) 3-core\end{tabular}}}}%
  \end{picture}%
\endgroup%

%% file: Figures/case5.pdf_tex
\begingroup%
  \makeatletter%
  \providecommand\color[2][]{%
    \errmessage{(Inkscape) Color is used for the text in Inkscape, but the package 'color.sty' is not loaded}%
    \renewcommand\color[2][]{}%
  }%
  \providecommand\transparent[1]{%
    \errmessage{(Inkscape) Transparency is used (non-zero) for the text in Inkscape, but the package 'transparent.sty' is not loaded}%
    \renewcommand\transparent[1]{}%
  }%
  \providecommand\rotatebox[2]{#2}%
  \newcommand*\fsize{\dimexpr\f@size pt\relax}%
  \newcommand*\lineheight[1]{\fontsize{\fsize}{#1\fsize}\selectfont}%
  \ifx\svgwidth\undefined%
    \setlength{\unitlength}{375bp}%
    \ifx\svgscale\undefined%
      \relax%
    \else%
      \setlength{\unitlength}{\unitlength * \real{\svgscale}}%
    \fi%
  \else%
    \setlength{\unitlength}{\svgwidth}%
  \fi%
  \global\let\svgwidth\undefined%
  \global\let\svgscale\undefined%
  \makeatother%
  \begin{picture}(1,1.1)%
    \lineheight{1}%
    \setlength\tabcolsep{0pt}%
    \put(0,0){\includegraphics[width=\unitlength,page=1]{case5.pdf}}%
    \put(0.69068977,0.88421552){\color[rgb]{0,0,0}\makebox(0,0)[lt]{\lineheight{1.25}\smash{\begin{tabular}[t]{l}C,D\end{tabular}}}}%
    \put(0,0){\includegraphics[width=\unitlength,page=2]{case5.pdf}}%
    \put(0.81868979,0.4842155){\color[rgb]{0,0,0}\makebox(0,0)[lt]{\lineheight{1.25}\smash{\begin{tabular}[t]{l}C,D\end{tabular}}}}%
    \put(0,0){\includegraphics[width=\unitlength,page=3]{case5.pdf}}%
    \put(0.49080392,0.2702155){\color[rgb]{0,0,0}\makebox(0,0)[t]{\lineheight{1.25}\smash{\begin{tabular}[t]{c}A,E\end{tabular}}}}%
    \put(0,0){\includegraphics[width=\unitlength,page=4]{case5.pdf}}%
    \put(0.16572136,0.52){\color[rgb]{0,0,0}\makebox(0,0)[lt]{\lineheight{1.25}\smash{\begin{tabular}[t]{l}B\end{tabular}}}}%
    \put(0,0){\includegraphics[width=\unitlength,page=5]{case5.pdf}}%
    \put(0.37970378,0.64421549){\color[rgb]{0,0,0}\makebox(0,0)[lt]{\lineheight{1.25}\smash{\begin{tabular}[t]{l}A,B,E\end{tabular}}}}%
    \put(0,0){\includegraphics[width=\unitlength,page=6]{case5.pdf}}%
    \put(0.48744376,1.00833333){\color[rgb]{1,1,1}\makebox(0,0)[t]{\lineheight{1.25}\smash{\begin{tabular}[t]{c}0\end{tabular}}}}%
    \put(0.52944376,0.99705554){\color[rgb]{0,0,0}\makebox(0,0)[lt]{\lineheight{1.25}\smash{\begin{tabular}[t]{l}Target: A\end{tabular}}}}%
    \put(0.52944376,1.02738889){\color[rgb]{0,0,0}\makebox(0,0)[lt]{\lineheight{1.25}\smash{\begin{tabular}[t]{l}Source: B,C,D,E\end{tabular}}}}%
    \put(0,0){\includegraphics[width=\unitlength,page=7]{case5.pdf}}%
    \put(0.88688751,0.71812045){\color[rgb]{0,0,0}\makebox(0,0)[t]{\lineheight{1.25}\smash{\begin{tabular}[t]{c}1\end{tabular}}}}%
    \put(0,0){\includegraphics[width=\unitlength,page=8]{case5.pdf}}%
    \put(0.73431355,0.2485462){\color[rgb]{1,1,1}\makebox(0,0)[t]{\lineheight{1.25}\smash{\begin{tabular}[t]{c}2\end{tabular}}}}%
    \put(0.77631355,0.23726842){\color[rgb]{0,0,0}\makebox(0,0)[lt]{\lineheight{1.25}\smash{\begin{tabular}[t]{l}Target: C,D,E\end{tabular}}}}%
    \put(0.77631355,0.26760176){\color[rgb]{0,0,0}\makebox(0,0)[lt]{\lineheight{1.25}\smash{\begin{tabular}[t]{l}Quelle: A\end{tabular}}}}%
    \put(0,0){\includegraphics[width=\unitlength,page=9]{case5.pdf}}%
    \put(0.24057396,0.2485462){\color[rgb]{1,1,1}\makebox(0,0)[t]{\lineheight{1.25}\smash{\begin{tabular}[t]{c}3\end{tabular}}}}%
    \put(0,0){\includegraphics[width=\unitlength,page=10]{case5.pdf}}%
    \put(0.088,0.71812045){\color[rgb]{1,1,1}\makebox(0,0)[t]{\lineheight{1.25}\smash{\begin{tabular}[t]{c}4\end{tabular}}}}%
    \put(0.13000002,0.70684269){\color[rgb]{0,0,0}\makebox(0,0)[lt]{\lineheight{1.25}\smash{\begin{tabular}[t]{l}Target: B\end{tabular}}}}%
    \put(0.015,0.13735801){\color[rgb]{0,0,0}\makebox(0,0)[lt]{\lineheight{1.25}\smash{\begin{tabular}[t]{l}Switch\end{tabular}}}}%
    \put(0,0){\includegraphics[width=\unitlength,page=11]{case5.pdf}}%
    \put(0.0612,0.09675801){\color[rgb]{0,0,0}\makebox(0,0)[lt]{\lineheight{1.25}\smash{\begin{tabular}[t]{l}(\#0) opaque\end{tabular}}}}%
    \put(0,0){\includegraphics[width=\unitlength,page=12]{case5.pdf}}%
    \put(0.0612,0.05475801){\color[rgb]{0,0,0}\makebox(0,0)[lt]{\lineheight{1.25}\smash{\begin{tabular}[t]{l}(\#1) transparent\end{tabular}}}}%
    \put(0.4,0.13735938){\color[rgb]{0,0,0}\makebox(0,0)[lt]{\lineheight{1.25}\smash{\begin{tabular}[t]{l}Cable\end{tabular}}}}%
    \put(0,0){\includegraphics[width=\unitlength,page=13]{case5.pdf}}%
    \put(0.44619999,0.09675938){\color[rgb]{0,0,0}\makebox(0,0)[lt]{\lineheight{1.25}\smash{\begin{tabular}[t]{l}(\#0) 1-core high attenuation\end{tabular}}}}%
    \put(0,0){\includegraphics[width=\unitlength,page=14]{case5.pdf}}%
    \put(0.44619999,0.05475938){\color[rgb]{0,0,0}\makebox(0,0)[lt]{\lineheight{1.25}\smash{\begin{tabular}[t]{l}(\#1) 2-core\end{tabular}}}}%
    \put(0,0){\includegraphics[width=\unitlength,page=15]{case5.pdf}}%
    \put(0.44619999,0.01275938){\color[rgb]{0,0,0}\makebox(0,0)[lt]{\lineheight{1.25}\smash{\begin{tabular}[t]{l}(\#2) 3-core\end{tabular}}}}%
  \end{picture}%
\endgroup%

%% file: Figures/case4f.pdf_tex
\begingroup%
  \makeatletter%
  \providecommand\color[2][]{%
    \errmessage{(Inkscape) Color is used for the text in Inkscape, but the package 'color.sty' is not loaded}%
    \renewcommand\color[2][]{}%
  }%
  \providecommand\transparent[1]{%
    \errmessage{(Inkscape) Transparency is used (non-zero) for the text in Inkscape, but the package 'transparent.sty' is not loaded}%
    \renewcommand\transparent[1]{}%
  }%
  \providecommand\rotatebox[2]{#2}%
  \newcommand*\fsize{\dimexpr\f@size pt\relax}%
  \newcommand*\lineheight[1]{\fontsize{\fsize}{#1\fsize}\selectfont}%
  \ifx\svgwidth\undefined%
    \setlength{\unitlength}{375bp}%
    \ifx\svgscale\undefined%
      \relax%
    \else%
      \setlength{\unitlength}{\unitlength * \real{\svgscale}}%
    \fi%
  \else%
    \setlength{\unitlength}{\svgwidth}%
  \fi%
  \global\let\svgwidth\undefined%
  \global\let\svgscale\undefined%
  \makeatother%
  \begin{picture}(1,1.1)%
    \lineheight{1}%
    \setlength\tabcolsep{0pt}%
    \put(0,0){\includegraphics[width=\unitlength,page=1]{case4f.pdf}}%
    \put(0.61653409,0.64422178){\color[rgb]{0,0,0}\makebox(0,0)[lt]{\lineheight{1.25}\smash{\begin{tabular}[t]{l}A,C,D\end{tabular}}}}%
    \put(0,0){\includegraphics[width=\unitlength,page=2]{case4f.pdf}}%
    \put(0.28715909,0.89254755){\color[rgb]{0,0,0}\makebox(0,0)[lt]{\lineheight{1.25}\smash{\begin{tabular}[t]{l}B\end{tabular}}}}%
    \put(0,0){\includegraphics[width=\unitlength,page=3]{case4f.pdf}}%
    \put(0.48744375,1.00833333){\color[rgb]{1,1,1}\makebox(0,0)[t]{\lineheight{1.25}\smash{\begin{tabular}[t]{c}0\end{tabular}}}}%
    \put(0.52944376,0.99705554){\color[rgb]{0,0,0}\makebox(0,0)[lt]{\lineheight{1.25}\smash{\begin{tabular}[t]{l}Target: A\end{tabular}}}}%
    \put(0.52944376,1.02738889){\color[rgb]{0,0,0}\makebox(0,0)[lt]{\lineheight{1.25}\smash{\begin{tabular}[t]{l}Source: B,C,D\end{tabular}}}}%
    \put(0,0){\includegraphics[width=\unitlength,page=4]{case4f.pdf}}%
    \put(0.88688752,0.71812048){\color[rgb]{0,0,0}\makebox(0,0)[t]{\lineheight{1.25}\smash{\begin{tabular}[t]{c}1\end{tabular}}}}%
    \put(0,0){\includegraphics[width=\unitlength,page=5]{case4f.pdf}}%
    \put(0.73431355,0.2485462){\color[rgb]{1,1,1}\makebox(0,0)[t]{\lineheight{1.25}\smash{\begin{tabular}[t]{c}2\end{tabular}}}}%
    \put(0.77631355,0.23726842){\color[rgb]{0,0,0}\makebox(0,0)[lt]{\lineheight{1.25}\smash{\begin{tabular}[t]{l}Target: C,D\end{tabular}}}}%
    \put(0.77631355,0.26760176){\color[rgb]{0,0,0}\makebox(0,0)[lt]{\lineheight{1.25}\smash{\begin{tabular}[t]{l}Source: A\end{tabular}}}}%
    \put(0,0){\includegraphics[width=\unitlength,page=6]{case4f.pdf}}%
    \put(0.24057397,0.2485462){\color[rgb]{0,0,0}\makebox(0,0)[t]{\lineheight{1.25}\smash{\begin{tabular}[t]{c}3\end{tabular}}}}%
    \put(0,0){\includegraphics[width=\unitlength,page=7]{case4f.pdf}}%
    \put(0.08799998,0.71812048){\color[rgb]{1,1,1}\makebox(0,0)[t]{\lineheight{1.25}\smash{\begin{tabular}[t]{c}4\end{tabular}}}}%
    \put(0.14079977,0.70094923){\color[rgb]{0,0,0}\makebox(0,0)[lt]{\lineheight{1.25}\smash{\begin{tabular}[t]{l}Target: B\end{tabular}}}}%
    \put(0.015,0.137358){\color[rgb]{0,0,0}\makebox(0,0)[lt]{\lineheight{1.25}\smash{\begin{tabular}[t]{l}Switch\end{tabular}}}}%
    \put(0,0){\includegraphics[width=\unitlength,page=8]{case4f.pdf}}%
    \put(0.0612,0.096758){\color[rgb]{0,0,0}\makebox(0,0)[lt]{\lineheight{1.25}\smash{\begin{tabular}[t]{l}(\#0) opaque\end{tabular}}}}%
    \put(0,0){\includegraphics[width=\unitlength,page=9]{case4f.pdf}}%
    \put(0.0612,0.054758){\color[rgb]{0,0,0}\makebox(0,0)[lt]{\lineheight{1.25}\smash{\begin{tabular}[t]{l}(\#1) transparent\end{tabular}}}}%
    \put(0.40000002,0.13735938){\color[rgb]{0,0,0}\makebox(0,0)[lt]{\lineheight{1.25}\smash{\begin{tabular}[t]{l}Cable\end{tabular}}}}%
    \put(0,0){\includegraphics[width=\unitlength,page=10]{case4f.pdf}}%
    \put(0.44620002,0.09675938){\color[rgb]{0,0,0}\makebox(0,0)[lt]{\lineheight{1.25}\smash{\begin{tabular}[t]{l}(\#0) 1-core high attenuation\end{tabular}}}}%
    \put(0,0){\includegraphics[width=\unitlength,page=11]{case4f.pdf}}%
    \put(0.44620002,0.05475938){\color[rgb]{0,0,0}\makebox(0,0)[lt]{\lineheight{1.25}\smash{\begin{tabular}[t]{l}(\#1) 2-core\end{tabular}}}}%
    \put(0,0){\includegraphics[width=\unitlength,page=12]{case4f.pdf}}%
    \put(0.44620002,0.01275938){\color[rgb]{0,0,0}\makebox(0,0)[lt]{\lineheight{1.25}\smash{\begin{tabular}[t]{l}(\#2) 3-core\end{tabular}}}}%
  \end{picture}%
\endgroup%

%% file: results.tex
	\section{Optimization of an In-flight Entertainment Network}
	\label{sec:results}

	An In-flight Entertainment system (IFE) scenario is used to demonstrate our approach on a larger scale. 

	The scenario comprises three rows of seats of an aircraft with a center aisle. Each row of seats consists of two groups of seats, one to the left and one to the right of the aisle. Each of these seating groups contains a central computer, called a \textit{Seat Unit}, which controls the screens and controls. There is a panel above each seating group with call buttons, indicator, and reading lights. Each panel has a dedicated control unit, i.e. \textit{Call and Light Panel}.
	The IFE is able to display live TV and movies at each seating area. Two streaming computers (\textit{Stream Server}) provide the data to be transmitted over the optical network. Both \textit{Stream Servers} can provide each seating group with a video stream. The \textit{Stream Servers} also control the lamps of the \textit{Call and Light Panels}.
	All inputs to the control elements of the seating groups and the call buttons are processed by two  \textit{Control Servers}. These command also the \textit{Stream Servers}. 
	The \textit{Control Servers} and \textit{Stream Servers} are fully bidirectionally networked for communication with each other.
	In a real aircraft, multiple of these aisle segments would be placed sequentially.

	\begin{figure}[htbp] 
		\centering
		\small
		\def\svgwidth{0.9\columnwidth}
		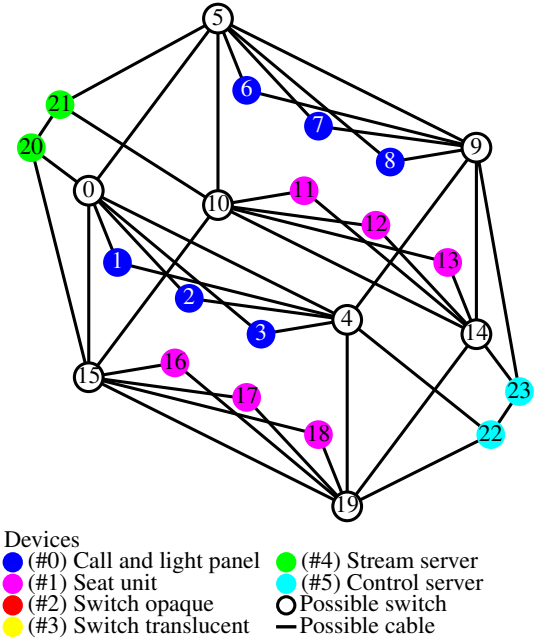
		\caption{Default network of the scenario IFE.}
		\label{fig:scn:ife2:pre}
	\end{figure}

The optimization space for the IFE scenario is shown in fig.\ref{fig:scn:ife2:pre}. The points outlined in black represent possible positions for switches. There are two types to choose from, namely an opaque switch and a translucent switch. All other devices have their type already assigned. The ID of devices are listed in table \ref{tbl:ife:devices}.

	\begin{table}[htbp]
		\centering
		\begin{tabular}{@{}ll@{}}
			\toprule
			\textbf{ID} & \textbf{Type} \\
			\midrule
			1, 2, 3  & Call and Light Panel (A)\\
			6, 7, 8  & Call and Light Panel (B)\\
			11, 12, 13 & Seat Unit (B)\\
			16, 17, 18 & Seat Unit (A)\\
			20, 21 & Stream Server\\
			22, 23 & Control Server\\
			
			0, 4, 5, 9, 10, 14, 15, 19 & possible Switches\\
			\bottomrule
		\end{tabular}
		\caption{Device and IDs used in the IFE scenario}
		\label{tbl:ife:devices}
	\end{table}

	Table \ref{tbl:ife:deviceTypes} lists the available device types with their properties. With the exception of one translucent switch, all devices are opaque.

	\begin{table}[htbp]
		\centering
		\setlength{\tabcolsep}{2pt}
		\scriptsize
		\begin{tabular}{@{}lcccccccc@{}}
			\toprule
			& \multicolumn{8}{c}{\textbf{Property}} \\
			\textbf{Name}
			& $\vec{\tau}{\set D}{\text{ports}_t}$
			& $\vec{\tau}{\set D}{\Delta_t}$
			& $\vec{\tau}{\set D}{\underlineText{Rx}_t}$
			& $\vec{\tau}{\set D}{\overlineText{Rx}_t}$
			& $\vec{\tau}{\set D}{\underlineText{Tx}_t}$
			& $\vec{\tau}{\set D}{\overlineText{Tx}_t}$
			& $\vec{\tau}{\set D}{\text{trans}_t}$
			& $\vec{\tau}{\set D}{\text{cost}_t}$
			\\
			\midrule
			
			Call and Light P.
				& \num{2}
				& ---
				& \SI{-14}
				& \SI{0.5}
				& \SI{-5}
				& \SI{0}
				& \num{0}
				& ---
			\\		
			
			Seat Unit
				& \num{2}
				& ---
				& \SI{-14}
				& \SI{0.5}
				& \SI{-5}
				& \SI{0}
				& \num{0}
				& ---
			\\

			Switch Opaque
				& \num{6}
				& ---
				& \SI{-14}
				& \SI{0.5}
				& \SI{-5}
				& \SI{0}
				& \num{0}
				& \num{6000}
			\\

			Switch Transl.
				& \num{6}
				& \SI{-0.5}{\decibel}
				& ---
				& ---
				& ---
				& ---
				& \num{1}
				& \num{5600}
			\\

			Stream Server
				& \num{3}
				& ---
				& \SI{-14}
				& \SI{0.5}
				& \SI{-5}
				& \SI{0}
				& \num{0}
				& ---
			\\

			Control Server
				& \num{3}
				& ---
				& \SI{-14}
				& \SI{0.5}
				& \SI{-5}
				& \SI{0}
				& \num{0}
				& ---
			\\
				
			\bottomrule
		\end{tabular}
		
		\caption{Device types and properties of IFE scenario. RX and TX entries are in {\dBm}. Entries marked with --- are modeled with the value zero.}
		\label{tbl:ife:deviceTypes}
	\end{table}

	Due to the fixed assignment of the device types \#0, \#1, \#4 and \#5, no cost values are used. Costs are assigned to the two switch types subject to the optimization. It is assumed that a translucent device is less expensive than an opaque device that requires active components.
	The available cable types are listed in table \ref{tbl:ife:cableTypes}. They all have identical attenuation characteristics, but differ based on their number of cores. The chosen numbers are inspired by the \cite{ds:cableLapp}. 
	The optimization objective is the total cost for cables and switches. 

	\begin{table}[htbp]
		\caption{Properties of cable types in IFE scenario. All cable types declared are bidirectional. Attenuation includes connectors and internal cable losses.}
		\centering
		\begin{tabular}{@{}llccc@{}}
			\toprule
			& & \multicolumn{3}{c}{\textbf{Property}} \\
			\textbf{Nb.} & \textbf{Name}
			& $\vec{\tau}{\set F}{\text{cores}_t}$
			& $\vec{\tau}{\set F}{\Delta_t}$
			& $\vec{\tau}{\set F}{\text{cost}_t}$
			\\
			\midrule
			
			\#0 & Optical Wire 2 Core
				& \num{2}
				& \SI{-2}{\decibel}
				& \num{10}
			\\

			\#1 & Optical Wire 4 Core
				& \num{4}
				& \SI{-2}{\decibel}
				& \num{40}
			\\

			\#2 & Optical Wire 8 Core
				& \num{8}
				& \SI{-2}{\decibel}
				& \num{80}
			\\

			\#3 & Optical Wire 12 Core
				& \num{12}
				& \SI{-2}{\decibel}
				& \num{90}
			\\
			
			\bottomrule
		\end{tabular}

		\label{tbl:ife:cableTypes}
	\end{table}

	With two exceptions, the cables are free of choice. Neither a type is assigned, nor is the selection restricted. Only the connection between the two \textit{Stream Servers} 20 and 21, as well as the connection between the two \textit{Control Servers} 22 and 23 are limited to the two-core cable type. 
	In total, \num{48} signals are necessary for this scenario as listed in table \ref{tbl:ife:signals}.

	\begin{table}[htbp]
		\setlength{\tabcolsep}{4pt}
		\begin{multicols}{2}
			\begin{tabularx}{\linewidth}{@{}cc>{\centering\arraybackslash}Xc@{}}
				\toprule
				\textbf{ID} & \textbf{S.} &	\textbf{Path} & \textbf{T.}  \\
				\midrule
				A  & 20 & -15- & 16 \\
				B  & 20 & -15- & 17 \\
				C  & 20 & -15- & 18 \\
				D  & 20 & -15-10- & 11 \\
				E  & 20 & -15-10- & 12 \\
				F  & 20 & -15-10- & 13 \\
				G  & 21 & -10-15- & 16 \\
				H  & 21 & -10-15- & 17 \\
				I  & 21 & -10-15- & 18 \\
				J  & 21 & -10-14- & 11 \\
				K  & 21 & -10- & 12 \\
				L  & 21 & -10- & 13 \\
				M  & 20 & -0-& 1 \\
				N  & 20 & -0-& 2 \\
				O  & 20 & -0-& 3 \\
				P  & 20 & -0-5- & 6 \\
				Q  & 20 & -0-5- & 7 \\
				R  & 20 & -0-5- & 8 \\
				S  & 21 & -5-0- & 1 \\
				T  & 21 & -5-0- & 2 \\
				U  & 21 & -5-0- & 3 \\
				V  & 21 & -5- & 6 \\
				W  & 21 & -5- & 7 \\
				X  & 21 & -5- & 8 \\
				Y  & 16 & -15-20-0-4- & 22 \\
				Z  & 17 & -15-20-0-4- & 22 \\
				
				\bottomrule
			\end{tabularx}
			
			\begin{tabularx}{\linewidth}{@{}cc>{\centering\arraybackslash}Xc@{}}
				\toprule
				\textbf{ID} & \textbf{S.} &	\textbf{Path} & \textbf{T.}  \\
				\midrule
				
				BA & 18 & -15-20-0-4- & 22 \\
				BB & 11 & -14- & 23 \\
				BC & 12 & -14- & 23 \\
				BD & 13 & -14- & 23 \\
				BE & 1  & -4- & 22 \\
				BF & 2  & -4- & 22 \\
				BG & 3  & -4- & 22 \\
				BH & 6  & -5-21-10-14- & 23 \\
				BI & 7  & -5-21-10-14- & 23 \\
				BJ & 8  & -5-21-10-14- & 23 \\
				BK & 22 & - & 23 \\
				BL & 23 & - & 22 \\
				BM & 20 & - & 21 \\
				BN & 21 & - & 20 \\
				BO & 22 & -4-0- & 20 \\
				BP & 22 & -4-0-5- & 21 \\
				BQ & 23 & -14-10-15- & 20 \\
				BR & 23 & -14-10- & 21 \\
				BS & 20 & -0-4- & 22 \\
				BT & 20 & -15-10-14- & 23 \\
				BU & 21 & -5-0-4- & 22 \\
				BV & 21 & -10-14- & 23 \\
				
				\bottomrule
			\end{tabularx}
			
		\end{multicols}
			
		\caption{Signal definitions and optimized paths for the IFE scenario. (S = source, T = target)}
		\label{tbl:ife:signals}
	\end{table}

	The IFE scenario resulted in a MILP with \num{49316} Constraints. \num{8454} binary, \num{162} integer and \num{9260} real variables are used.
	Finding a first solution requires \SI{12}{\second}. After three minutes, the solution within a limit of one percent deviation from the global optimum is obtained. It takes another \SI{9}{\minute} to confirm an optimal solution. The entire program execution requires \SI{23}{\minute}, of which a total of \SI{6}{\minute} are spent on data input, output, and internal model building. Another \SI{6}{\minute} are attributable to Gurobi and subsequent processing. 

\begin{figure}[htbp]
		\centering
		\small
		\def\svgwidth{0.9\columnwidth}
		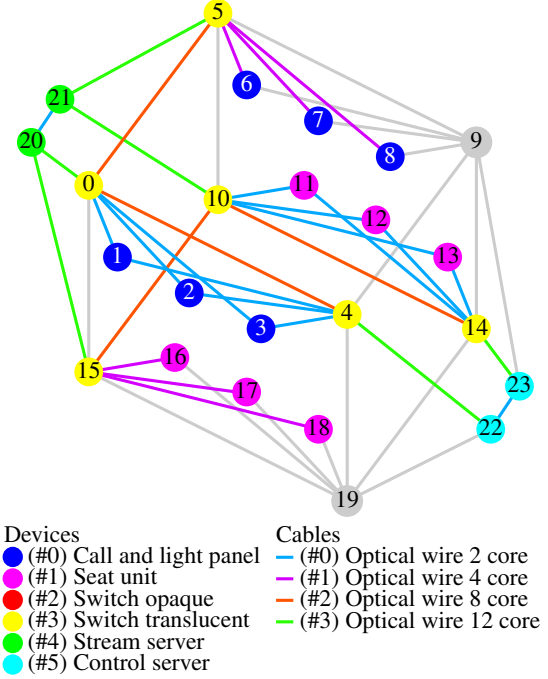
		\caption{Spatial representation of the optimal solution of the IFE scenario. The colors represent the assigned device and cable types. Gray devices and cables were removed by the optimizer.}
		\label{fig:scn:ife2:result3D}
	\end{figure}

	The optimized topology is shown in fig. \ref{fig:scn:ife2:result3D}. 
	Six of the eight possible switches are instantiated in the topology as translucent devices (yellow). The remaining two possible switch positions are unused.
	The structure of the cables is interesting. The cable assignment is symmetrical along one axis. Although there is a symmetry of the defined signals in in the top strands (0-4 and 5-9) as well as in the bottom strands (10-14 and 15-19), this symmetry is not reflected in the optimized topology. Instead, a symmetry of left and right is observed, resulting from the control and stream device positions.

	A careful inspection of the optimization result uncovers a shortcoming in our cost function: While two signals to the Seat Unit with number 11 follow short paths from start to destination, one signal is not optimal in terms of hop count. Instead of using the available core of the connection 10-11, it runs via Switch~14. However, since the alternative path with fewer hops does not result in a better objective function value, it is not picked by the optimizer.

%% file: Figures/preOptimierung3D.pdf_tex
\begingroup%
  \makeatletter%
  \providecommand\color[2][]{%
    \errmessage{(Inkscape) Color is used for the text in Inkscape, but the package 'color.sty' is not loaded}%
    \renewcommand\color[2][]{}%
  }%
  \providecommand\transparent[1]{%
    \errmessage{(Inkscape) Transparency is used (non-zero) for the text in Inkscape, but the package 'transparent.sty' is not loaded}%
    \renewcommand\transparent[1]{}%
  }%
  \providecommand\rotatebox[2]{#2}%
  \newcommand*\fsize{\dimexpr\f@size pt\relax}%
  \newcommand*\lineheight[1]{\fontsize{\fsize}{#1\fsize}\selectfont}%
  \ifx\svgwidth\undefined%
    \setlength{\unitlength}{400bp}%
    \ifx\svgscale\undefined%
      \relax%
    \else%
      \setlength{\unitlength}{\unitlength * \real{\svgscale}}%
    \fi%
  \else%
    \setlength{\unitlength}{\svgwidth}%
  \fi%
  \global\let\svgwidth\undefined%
  \global\let\svgscale\undefined%
  \makeatother%
  \begin{picture}(1,1.17125)%
    \lineheight{1}%
    \setlength\tabcolsep{0pt}%
    \put(0,0){\includegraphics[width=\unitlength,page=1]{preOptimierung3D.pdf}}%
    \put(0.15750002,0.46437501){\color[rgb]{0,0,0}\makebox(0,0)[t]{\lineheight{1.25}\smash{\begin{tabular}[t]{c}15\end{tabular}}}}%
    \put(0,0){\includegraphics[width=\unitlength,page=2]{preOptimierung3D.pdf}}%
    \put(0.31500004,0.49062509){\color[rgb]{0,0,0}\makebox(0,0)[t]{\lineheight{1.25}\smash{\begin{tabular}[t]{c}16\end{tabular}}}}%
    \put(0,0){\includegraphics[width=\unitlength,page=3]{preOptimierung3D.pdf}}%
    \put(0.44625005,0.42749996){\color[rgb]{0,0,0}\makebox(0,0)[t]{\lineheight{1.25}\smash{\begin{tabular}[t]{c}17\end{tabular}}}}%
    \put(0,0){\includegraphics[width=\unitlength,page=4]{preOptimierung3D.pdf}}%
    \put(0.57750007,0.35937508){\color[rgb]{0,0,0}\makebox(0,0)[t]{\lineheight{1.25}\smash{\begin{tabular}[t]{c}18\end{tabular}}}}%
    \put(0,0){\includegraphics[width=\unitlength,page=5]{preOptimierung3D.pdf}}%
    \put(0.6300001,0.22812513){\color[rgb]{0,0,0}\makebox(0,0)[t]{\lineheight{1.25}\smash{\begin{tabular}[t]{c}19\end{tabular}}}}%
    \put(0,0){\includegraphics[width=\unitlength,page=6]{preOptimierung3D.pdf}}%
    \put(0.05250001,0.88437514){\color[rgb]{0,0,0}\makebox(0,0)[t]{\lineheight{1.25}\smash{\begin{tabular}[t]{c}20\end{tabular}}}}%
    \put(0,0){\includegraphics[width=\unitlength,page=7]{preOptimierung3D.pdf}}%
    \put(0.105,0.96312515){\color[rgb]{0,0,0}\makebox(0,0)[t]{\lineheight{1.25}\smash{\begin{tabular}[t]{c}21\end{tabular}}}}%
    \put(0,0){\includegraphics[width=\unitlength,page=8]{preOptimierung3D.pdf}}%
    \put(0.89250008,0.35937509){\color[rgb]{0,0,0}\makebox(0,0)[t]{\lineheight{1.25}\smash{\begin{tabular}[t]{c}22\end{tabular}}}}%
    \put(0,0){\includegraphics[width=\unitlength,page=9]{preOptimierung3D.pdf}}%
    \put(0.94500009,0.43812508){\color[rgb]{0,0,0}\makebox(0,0)[t]{\lineheight{1.25}\smash{\begin{tabular}[t]{c}23\end{tabular}}}}%
    \put(0,0){\includegraphics[width=\unitlength,page=10]{preOptimierung3D.pdf}}%
    \put(0.15750004,0.80562511){\color[rgb]{0,0,0}\makebox(0,0)[t]{\lineheight{1.25}\smash{\begin{tabular}[t]{c}0\end{tabular}}}}%
    \put(0,0){\includegraphics[width=\unitlength,page=11]{preOptimierung3D.pdf}}%
    \put(0.21000008,0.67437514){\color[rgb]{1,1,1}\makebox(0,0)[t]{\lineheight{1.25}\smash{\begin{tabular}[t]{c}1\end{tabular}}}}%
    \put(0,0){\includegraphics[width=\unitlength,page=12]{preOptimierung3D.pdf}}%
    \put(0.34125004,0.60875014){\color[rgb]{1,1,1}\makebox(0,0)[t]{\lineheight{1.25}\smash{\begin{tabular}[t]{c}2\end{tabular}}}}%
    \put(0,0){\includegraphics[width=\unitlength,page=13]{preOptimierung3D.pdf}}%
    \put(0.47250003,0.54312513){\color[rgb]{1,1,1}\makebox(0,0)[t]{\lineheight{1.25}\smash{\begin{tabular}[t]{c}3\end{tabular}}}}%
    \put(0,0){\includegraphics[width=\unitlength,page=14]{preOptimierung3D.pdf}}%
    \put(0.63000004,0.56937513){\color[rgb]{0,0,0}\makebox(0,0)[t]{\lineheight{1.25}\smash{\begin{tabular}[t]{c}4\end{tabular}}}}%
    \put(0,0){\includegraphics[width=\unitlength,page=15]{preOptimierung3D.pdf}}%
    \put(0.39374994,0.77937503){\color[rgb]{0,0,0}\makebox(0,0)[t]{\lineheight{1.25}\smash{\begin{tabular}[t]{c}10\end{tabular}}}}%
    \put(0,0){\includegraphics[width=\unitlength,page=16]{preOptimierung3D.pdf}}%
    \put(0.55124999,0.80562513){\color[rgb]{0,0,0}\makebox(0,0)[t]{\lineheight{1.25}\smash{\begin{tabular}[t]{c}11\end{tabular}}}}%
    \put(0,0){\includegraphics[width=\unitlength,page=17]{preOptimierung3D.pdf}}%
    \put(0.68250003,0.74250012){\color[rgb]{0,0,0}\makebox(0,0)[t]{\lineheight{1.25}\smash{\begin{tabular}[t]{c}12\end{tabular}}}}%
    \put(0,0){\includegraphics[width=\unitlength,page=18]{preOptimierung3D.pdf}}%
    \put(0.81375003,0.67437507){\color[rgb]{0,0,0}\makebox(0,0)[t]{\lineheight{1.25}\smash{\begin{tabular}[t]{c}13\end{tabular}}}}%
    \put(0,0){\includegraphics[width=\unitlength,page=19]{preOptimierung3D.pdf}}%
    \put(0.86625014,0.54312504){\color[rgb]{0,0,0}\makebox(0,0)[t]{\lineheight{1.25}\smash{\begin{tabular}[t]{c}14\end{tabular}}}}%
    \put(0,0){\includegraphics[width=\unitlength,page=20]{preOptimierung3D.pdf}}%
    \put(0.39375005,1.12062513){\color[rgb]{0,0,0}\makebox(0,0)[t]{\lineheight{1.25}\smash{\begin{tabular}[t]{c}5\end{tabular}}}}%
    \put(0,0){\includegraphics[width=\unitlength,page=21]{preOptimierung3D.pdf}}%
    \put(0.4462501,0.98937513){\color[rgb]{1,1,1}\makebox(0,0)[t]{\lineheight{1.25}\smash{\begin{tabular}[t]{c}6\end{tabular}}}}%
    \put(0,0){\includegraphics[width=\unitlength,page=22]{preOptimierung3D.pdf}}%
    \put(0.57750007,0.92375005){\color[rgb]{1,1,1}\makebox(0,0)[t]{\lineheight{1.25}\smash{\begin{tabular}[t]{c}7\end{tabular}}}}%
    \put(0,0){\includegraphics[width=\unitlength,page=23]{preOptimierung3D.pdf}}%
    \put(0.70875011,0.85812512){\color[rgb]{1,1,1}\makebox(0,0)[t]{\lineheight{1.25}\smash{\begin{tabular}[t]{c}8\end{tabular}}}}%
    \put(0,0){\includegraphics[width=\unitlength,page=24]{preOptimierung3D.pdf}}%
    \put(0.86625014,0.88437504){\color[rgb]{0,0,0}\makebox(0,0)[t]{\lineheight{1.25}\smash{\begin{tabular}[t]{c}9\end{tabular}}}}%
    \put(0.0028125,0.16424989){\color[rgb]{0,0,0}\makebox(0,0)[lt]{\lineheight{1.25}\smash{\begin{tabular}[t]{l}Devices\end{tabular}}}}%
    \put(0,0){\includegraphics[width=\unitlength,page=25]{preOptimierung3D.pdf}}%
    \put(0.046125,0.12618745){\color[rgb]{0,0,0}\makebox(0,0)[lt]{\lineheight{1.25}\smash{\begin{tabular}[t]{l}(\#0) Call and light panel\end{tabular}}}}%
    \put(0,0){\includegraphics[width=\unitlength,page=26]{preOptimierung3D.pdf}}%
    \put(0.046125,0.08681248){\color[rgb]{0,0,0}\makebox(0,0)[lt]{\lineheight{1.25}\smash{\begin{tabular}[t]{l}(\#1) Seat unit\end{tabular}}}}%
    \put(0,0){\includegraphics[width=\unitlength,page=27]{preOptimierung3D.pdf}}%
    \put(0.046125,0.04743752){\color[rgb]{0,0,0}\makebox(0,0)[lt]{\lineheight{1.25}\smash{\begin{tabular}[t]{l}(\#2) Switch opaque\end{tabular}}}}%
    \put(0,0){\includegraphics[width=\unitlength,page=28]{preOptimierung3D.pdf}}%
    \put(0.046125,0.00806256){\color[rgb]{0,0,0}\makebox(0,0)[lt]{\lineheight{1.25}\smash{\begin{tabular}[t]{l}(\#3) Switch translucent\end{tabular}}}}%
    \put(0,0){\includegraphics[width=\unitlength,page=29]{preOptimierung3D.pdf}}%
    \put(0.54331251,0.008212){\color[rgb]{0,0,0}\makebox(0,0)[lt]{\lineheight{1.25}\smash{\begin{tabular}[t]{l}Possible cable\end{tabular}}}}%
    \put(0.54331251,0.04758696){\color[rgb]{0,0,0}\makebox(0,0)[lt]{\lineheight{1.25}\smash{\begin{tabular}[t]{l}Possible switch\end{tabular}}}}%
    \put(0,0){\includegraphics[width=\unitlength,page=30]{preOptimierung3D.pdf}}%
    \put(0.54612502,0.12628749){\color[rgb]{0,0,0}\makebox(0,0)[lt]{\lineheight{1.25}\smash{\begin{tabular}[t]{l}(\#4) Stream server\end{tabular}}}}%
    \put(0,0){\includegraphics[width=\unitlength,page=31]{preOptimierung3D.pdf}}%
    \put(0.54612502,0.08691253){\color[rgb]{0,0,0}\makebox(0,0)[lt]{\lineheight{1.25}\smash{\begin{tabular}[t]{l}(\#5) Control server\end{tabular}}}}%
  \end{picture}%
\endgroup%

%% file: Figures/3D.pdf_tex
\begingroup%
  \makeatletter%
  \providecommand\color[2][]{%
    \errmessage{(Inkscape) Color is used for the text in Inkscape, but the package 'color.sty' is not loaded}%
    \renewcommand\color[2][]{}%
  }%
  \providecommand\transparent[1]{%
    \errmessage{(Inkscape) Transparency is used (non-zero) for the text in Inkscape, but the package 'transparent.sty' is not loaded}%
    \renewcommand\transparent[1]{}%
  }%
  \providecommand\rotatebox[2]{#2}%
  \newcommand*\fsize{\dimexpr\f@size pt\relax}%
  \newcommand*\lineheight[1]{\fontsize{\fsize}{#1\fsize}\selectfont}%
  \ifx\svgwidth\undefined%
    \setlength{\unitlength}{400bp}%
    \ifx\svgscale\undefined%
      \relax%
    \else%
      \setlength{\unitlength}{\unitlength * \real{\svgscale}}%
    \fi%
  \else%
    \setlength{\unitlength}{\svgwidth}%
  \fi%
  \global\let\svgwidth\undefined%
  \global\let\svgscale\undefined%
  \makeatother%
  \begin{picture}(1,1.25)%
    \lineheight{1}%
    \setlength\tabcolsep{0pt}%
    \put(0,0){\includegraphics[width=\unitlength,page=1]{3D.pdf}}%
    \put(0.39375005,1.19937518){\color[rgb]{0,0,0}\makebox(0,0)[t]{\lineheight{1.25}\smash{\begin{tabular}[t]{c}5\end{tabular}}}}%
    \put(0,0){\includegraphics[width=\unitlength,page=2]{3D.pdf}}%
    \put(0.4462501,1.06812512){\color[rgb]{1,1,1}\makebox(0,0)[t]{\lineheight{1.25}\smash{\begin{tabular}[t]{c}6\end{tabular}}}}%
    \put(0,0){\includegraphics[width=\unitlength,page=3]{3D.pdf}}%
    \put(0.57750007,1.00250004){\color[rgb]{1,1,1}\makebox(0,0)[t]{\lineheight{1.25}\smash{\begin{tabular}[t]{c}7\end{tabular}}}}%
    \put(0,0){\includegraphics[width=\unitlength,page=4]{3D.pdf}}%
    \put(0.70875012,0.93687512){\color[rgb]{1,1,1}\makebox(0,0)[t]{\lineheight{1.25}\smash{\begin{tabular}[t]{c}8\end{tabular}}}}%
    \put(0,0){\includegraphics[width=\unitlength,page=5]{3D.pdf}}%
    \put(0.8662501,0.96312511){\color[rgb]{0,0,0}\makebox(0,0)[t]{\lineheight{1.25}\smash{\begin{tabular}[t]{c}9\end{tabular}}}}%
    \put(0,0){\includegraphics[width=\unitlength,page=6]{3D.pdf}}%
    \put(0.39374995,0.8581251){\color[rgb]{0,0,0}\makebox(0,0)[t]{\lineheight{1.25}\smash{\begin{tabular}[t]{c}10\end{tabular}}}}%
    \put(0,0){\includegraphics[width=\unitlength,page=7]{3D.pdf}}%
    \put(0.55124999,0.88437513){\color[rgb]{0,0,0}\makebox(0,0)[t]{\lineheight{1.25}\smash{\begin{tabular}[t]{c}11\end{tabular}}}}%
    \put(0,0){\includegraphics[width=\unitlength,page=8]{3D.pdf}}%
    \put(0.68250003,0.82125012){\color[rgb]{0,0,0}\makebox(0,0)[t]{\lineheight{1.25}\smash{\begin{tabular}[t]{c}12\end{tabular}}}}%
    \put(0,0){\includegraphics[width=\unitlength,page=9]{3D.pdf}}%
    \put(0.81375003,0.75312506){\color[rgb]{0,0,0}\makebox(0,0)[t]{\lineheight{1.25}\smash{\begin{tabular}[t]{c}13\end{tabular}}}}%
    \put(0,0){\includegraphics[width=\unitlength,page=10]{3D.pdf}}%
    \put(0.86625008,0.62187506){\color[rgb]{0,0,0}\makebox(0,0)[t]{\lineheight{1.25}\smash{\begin{tabular}[t]{c}14\end{tabular}}}}%
    \put(0,0){\includegraphics[width=\unitlength,page=11]{3D.pdf}}%
    \put(0.15750002,0.54312508){\color[rgb]{0,0,0}\makebox(0,0)[t]{\lineheight{1.25}\smash{\begin{tabular}[t]{c}15\end{tabular}}}}%
    \put(0,0){\includegraphics[width=\unitlength,page=12]{3D.pdf}}%
    \put(0.31500004,0.56937508){\color[rgb]{0,0,0}\makebox(0,0)[t]{\lineheight{1.25}\smash{\begin{tabular}[t]{c}16\end{tabular}}}}%
    \put(0,0){\includegraphics[width=\unitlength,page=13]{3D.pdf}}%
    \put(0.44625005,0.50624995){\color[rgb]{0,0,0}\makebox(0,0)[t]{\lineheight{1.25}\smash{\begin{tabular}[t]{c}17\end{tabular}}}}%
    \put(0,0){\includegraphics[width=\unitlength,page=14]{3D.pdf}}%
    \put(0.57750007,0.43812506){\color[rgb]{0,0,0}\makebox(0,0)[t]{\lineheight{1.25}\smash{\begin{tabular}[t]{c}18\end{tabular}}}}%
    \put(0,0){\includegraphics[width=\unitlength,page=15]{3D.pdf}}%
    \put(0.63000007,0.30687505){\color[rgb]{0,0,0}\makebox(0,0)[t]{\lineheight{1.25}\smash{\begin{tabular}[t]{c}19\end{tabular}}}}%
    \put(0,0){\includegraphics[width=\unitlength,page=16]{3D.pdf}}%
    \put(0.05250001,0.96312512){\color[rgb]{0,0,0}\makebox(0,0)[t]{\lineheight{1.25}\smash{\begin{tabular}[t]{c}20\end{tabular}}}}%
    \put(0,0){\includegraphics[width=\unitlength,page=17]{3D.pdf}}%
    \put(0.105,1.04187514){\color[rgb]{0,0,0}\makebox(0,0)[t]{\lineheight{1.25}\smash{\begin{tabular}[t]{c}21\end{tabular}}}}%
    \put(0,0){\includegraphics[width=\unitlength,page=18]{3D.pdf}}%
    \put(0.89250008,0.43812506){\color[rgb]{0,0,0}\makebox(0,0)[t]{\lineheight{1.25}\smash{\begin{tabular}[t]{c}22\end{tabular}}}}%
    \put(0,0){\includegraphics[width=\unitlength,page=19]{3D.pdf}}%
    \put(0.94500009,0.51687506){\color[rgb]{0,0,0}\makebox(0,0)[t]{\lineheight{1.25}\smash{\begin{tabular}[t]{c}23\end{tabular}}}}%
    \put(0,0){\includegraphics[width=\unitlength,page=20]{3D.pdf}}%
    \put(0.21000008,0.75312514){\color[rgb]{1,1,1}\makebox(0,0)[t]{\lineheight{1.25}\smash{\begin{tabular}[t]{c}1\end{tabular}}}}%
    \put(0,0){\includegraphics[width=\unitlength,page=21]{3D.pdf}}%
    \put(0.34125004,0.68750013){\color[rgb]{1,1,1}\makebox(0,0)[t]{\lineheight{1.25}\smash{\begin{tabular}[t]{c}2\end{tabular}}}}%
    \put(0,0){\includegraphics[width=\unitlength,page=22]{3D.pdf}}%
    \put(0.47250003,0.62187513){\color[rgb]{1,1,1}\makebox(0,0)[t]{\lineheight{1.25}\smash{\begin{tabular}[t]{c}3\end{tabular}}}}%
    \put(0,0){\includegraphics[width=\unitlength,page=23]{3D.pdf}}%
    \put(0.63000004,0.64812513){\color[rgb]{0,0,0}\makebox(0,0)[t]{\lineheight{1.25}\smash{\begin{tabular}[t]{c}4\end{tabular}}}}%
    \put(0,0){\includegraphics[width=\unitlength,page=24]{3D.pdf}}%
    \put(0.15750004,0.88437516){\color[rgb]{0,0,0}\makebox(0,0)[t]{\lineheight{1.25}\smash{\begin{tabular}[t]{c}0\end{tabular}}}}%
    \put(0.0028125,0.243){\color[rgb]{0,0,0}\makebox(0,0)[lt]{\lineheight{1.25}\smash{\begin{tabular}[t]{l}Devices\end{tabular}}}}%
    \put(0,0){\includegraphics[width=\unitlength,page=25]{3D.pdf}}%
    \put(0.046125,0.2049375){\color[rgb]{0,0,0}\makebox(0,0)[lt]{\lineheight{1.25}\smash{\begin{tabular}[t]{l}(\#0) Call and light panel\end{tabular}}}}%
    \put(0,0){\includegraphics[width=\unitlength,page=26]{3D.pdf}}%
    \put(0.046125,0.16556249){\color[rgb]{0,0,0}\makebox(0,0)[lt]{\lineheight{1.25}\smash{\begin{tabular}[t]{l}(\#1) Seat unit\end{tabular}}}}%
    \put(0,0){\includegraphics[width=\unitlength,page=27]{3D.pdf}}%
    \put(0.046125,0.12618749){\color[rgb]{0,0,0}\makebox(0,0)[lt]{\lineheight{1.25}\smash{\begin{tabular}[t]{l}(\#2) Switch opaque\end{tabular}}}}%
    \put(0,0){\includegraphics[width=\unitlength,page=28]{3D.pdf}}%
    \put(0.046125,0.08681249){\color[rgb]{0,0,0}\makebox(0,0)[lt]{\lineheight{1.25}\smash{\begin{tabular}[t]{l}(\#3) Switch translucent\end{tabular}}}}%
    \put(0,0){\includegraphics[width=\unitlength,page=29]{3D.pdf}}%
    \put(0.046125,0.04743749){\color[rgb]{0,0,0}\makebox(0,0)[lt]{\lineheight{1.25}\smash{\begin{tabular}[t]{l}(\#4) Stream server\end{tabular}}}}%
    \put(0,0){\includegraphics[width=\unitlength,page=30]{3D.pdf}}%
    \put(0.046125,0.00806248){\color[rgb]{0,0,0}\makebox(0,0)[lt]{\lineheight{1.25}\smash{\begin{tabular}[t]{l}(\#5) Control server\end{tabular}}}}%
    \put(0.5,0.24238474){\color[rgb]{0,0,0}\makebox(0,0)[lt]{\lineheight{1.25}\smash{\begin{tabular}[t]{l}Cables\end{tabular}}}}%
    \put(0,0){\includegraphics[width=\unitlength,page=31]{3D.pdf}}%
    \put(0.5433125,0.20432224){\color[rgb]{0,0,0}\makebox(0,0)[lt]{\lineheight{1.25}\smash{\begin{tabular}[t]{l}(\#0) Optical wire 2 core\end{tabular}}}}%
    \put(0,0){\includegraphics[width=\unitlength,page=32]{3D.pdf}}%
    \put(0.5433125,0.16494723){\color[rgb]{0,0,0}\makebox(0,0)[lt]{\lineheight{1.25}\smash{\begin{tabular}[t]{l}(\#1) Optical wire 4 core\end{tabular}}}}%
    \put(0,0){\includegraphics[width=\unitlength,page=33]{3D.pdf}}%
    \put(0.5433125,0.12557223){\color[rgb]{0,0,0}\makebox(0,0)[lt]{\lineheight{1.25}\smash{\begin{tabular}[t]{l}(\#2) Optical wire 8 core\end{tabular}}}}%
    \put(0,0){\includegraphics[width=\unitlength,page=34]{3D.pdf}}%
    \put(0.5433125,0.08619723){\color[rgb]{0,0,0}\makebox(0,0)[lt]{\lineheight{1.25}\smash{\begin{tabular}[t]{l}(\#3) Optical wire 12 core\end{tabular}}}}%
  \end{picture}%
\endgroup%

%% file: conclusion.tex
\section{Conclusion}
\label{sec:conclusion}
A topology optimization for an optimal network considering attenuation, multi-core fibers as well as translucent and opaque switches can be represented and solved as a MILP problem.
The validation scenarios indicate a correct functioning of the core-count, direction, and attenuation constraints.
The demonstration with the small IFE system shows that the methodology is appropriate for envisioned topology design. 30 min are an acceptable optimization time. 
However, further routing constraints (e.g. hop count and segregation) are needed to limit the signals routed through a device. 
The used domain-specific model OAAM includes ways to specify such constraints, which were not considered so far.
A critical consideration is the problem size. 
For the demonstrated system section, already a five-digit number of variables and constraints are necessary. Surprisingly, the demonstration model was solved comparatively quickly. 
Comparing the presented approach for optical networks with the optimization of classical networks, runtimes in the order of several hours to days can be expected for larger problems, e.g. a complete avionics system.
In future, the limitations imposed will be revised: 
The fact that only one single signal is transmitted in one core is a limitation in practice. Future implementations should, therefore, look at relaxing this limitation.
In addition, for future systems, it makes sense to extend the approach to other optical transmission techniques, primarily the commonly used Wavelength-Division-Multiplexing technique. 
Each technology has its own advantages and disadvantages, which is why there is also a potential need for optimization in the selection of a suitable technology.
In conclusion, it can be stated that topology optimization of optical networks is feasible within an acceptable computation time and provides added value to the design of aircraft systems.